\documentclass[twocolumn,secnumarabic,amssymb, nobibnotes, aps, prd]{revtex4}

\newcommand{\env}[1]{\texttt{#1}}
\usepackage{graphicx,epstopdf,color}
\usepackage{amsmath,amsthm,amssymb}

\def\fun#1#2{\lower3.6pt\vbox{\baselineskip0pt\lineskip.9pt
  \ialign{$\mathsurround=0pt#1\hfil##\hfil$\crcr#2\crcr\sim\crcr}}}

\newcommand{\be}{\begin{equation}}
\newcommand{\ee}{\end{equation}}
\newcommand{\ba}{\begin{eqnarray}}
\newcommand{\ea}{\end{eqnarray}}
\newcommand{\nn}{\nonumber}

\def\lsim{\raise0.3ex\hbox{$\;<$\kern-0.75em\raise-1.1ex\hbox{$\sim\;$}}}
\def\gsim{\raise0.3ex\hbox{$\;>$\kern-0.75em\raise-1.1ex\hbox{$\sim\;$}}}
\def\eps{\varepsilon}
\def\theta{\vartheta}

\renewcommand{\vec}[1]{\boldsymbol{#1}}

\begin{document}

\title{Anisotropic Cosmic Ray Diffusion and its Implications for Gamma-Ray Astronomy}

\author{G.~Giacinti$^{1}$}
\author{M.~Kachelrie\ss$^{2}$}
\author{D.~V.~Semikoz$^{3,4}$}
\affiliation{$^1$University of Oxford, Clarendon Laboratory, Oxford, United Kingdom}
\affiliation{$^2$Institutt for fysikk, NTNU, Trondheim, Norway}
\affiliation{$^3$AstroParticle and Cosmology (APC), Paris, France}
\affiliation{$^4$Institute for Nuclear Research of the Russian Academy of Sciences, Moscow, Russia}

\begin{abstract}
Analyses of TeV--PeV cosmic ray (CR) diffusion around their sources usually assume either isotropic diffusion or anisotropic diffusion due to the regular Galactic magnetic field. We show that none of them are adequate on distances smaller than the maximal scale $l_{\max}\sim 100$\,pc of fluctuations in the turbulent interstellar magnetic field. As a result, we predict anisotropic gamma-ray emissions around CR proton and electron sources, even for uniform densities of target gas. The centers of extended emission regions may have non-negligible offsets from their sources, leading to risks of misidentification. Gamma-rays from CR filaments have steeper energy spectra than those from surrounding regions. We point out that gamma-ray telescopes can be used in the future as a new way to probe and deduce the parameters of the interstellar magnetic field.
\end{abstract}

\date{\today}


\maketitle


\vspace{3pc}


\section{Introduction}
\label{Introduction}

Recent results on the anisotropy of cosmic rays (CRs) from the Pierre 
Auger Observatory~\cite{Abreu:2012lva} limit the dipole amplitude to 
$\lsim 1 \%$ at energies around 1\,EeV and thereby constrain the
contribution from Galactic sources to the dipole anisotropy to this level.
Also, the composition studies performed by the Kascade-GRANDE 
Collaboration~\cite{Apel:2013ura} show that the
fraction of the light component in the total spectrum rises for 
$E\gsim 10^{17}$\,eV. Combining these observations with studies of CR propagation 
in the Galactic magnetic field (GMF), one concludes that the transition 
from Galactic to extragalactic CRs happens between $10^{17}$\,eV and a 
few~$\times 10^{18}$\,eV~\cite{nuc3}.
A longstanding suggestion is that Galactic 
CRs are accelerated using the energy released in supernova (SN) explosions: 
Cosmic ray protons and electrons can be accelerated at the shock front of
SN remnants (SNR) or by electromagnetic induction in the magnetic field
of fast rotating pulsars. But since TeV--PeV CRs 
diffuse in the turbulent interstellar magnetic fields, directional information 
about their sources is absent from their arrival directions at Earth, and
thus a direct identification of their sources using CRs is impossible.

When CR protons encounter interstellar gas, they produce neutral pions, which 
in turn decay into gamma-rays. These gamma-rays offer a compelling way to 
detect the acceleration sites of CR protons, if they can be distinguished from
gamma-rays produced by electrons via bremsstrahlung and inverse Compton 
scattering. Such a differentiation can be accomplished by three means:
First, the characteristic feature that the photon flux $dN/d\ln(E)$ from  
pion-decay is symmetric with respect to $m_{\pi^0}/2$ and thus quickly 
decreasing for energies $<m_{\pi^0}/2$ can be employed~\cite{Ackermann:2013wqa}.
Second, the source and its parameters can be modelled and the relative
contribution from leptonic and hadronic processes to the photon flux
can be disentangled. Finally, one can test if the photon flux 
correlates with the gas density close to the source, i.e., test if molecular 
clouds are prominent photon sources~\cite{molcl}.

The last possibility requires an understanding of TeV CR diffusion on
scales smaller than or comparable with the maximal scale $l_{\max}$ of fluctuations 
in the turbulent GMF. Observations suggest that 
$l_{\max}={\cal O}(100\,{\rm pc})$~\cite{Han:2004aa}, which is at least two orders of magnitude 
larger than the Larmor radius $r_{\rm L}$ of PeV CRs, while the field is coherent on scales $l_{\rm c}$ of at least a few tens of 
parsecs~\cite{Han:2004aa,Frisch:2012zj,Frisch:2011qf,Frisch:2010fw}. 
One usually assumes that CRs diffuse either isotropically around their sources, or 
mildly anisotropically due to the regular Galactic magnetic field that elongates the CR distribution along its direction. 
However, we have shown in a recent study~\cite{Giacinti:2012ar} that these descriptions are incorrect on such scales: 
Since modes of the turbulent magnetic field $\vec B(\vec k)$ 
with scales $1/k\gg r_{\rm L}$ mimic a local regular field, diffusion proceeds strongly anisotropically even for an isotropic 
random field. 
For typical parameters of the GMF, this only locally uniform field dominates
over the regular component of the GMF.
As a result, CRs propagate preferentially in filamentary and twisted 
structures around young sources, which in turn can explain the irregular 
images of extended sources found, for instance, in Refs.~\cite{HESS,VERITAS,NS12}.
Moreover, the enhanced CR density in filamentary structures
may explain why the diffusion coefficient deduced by assuming isotropic 
diffusion is often below expectations close to CR sources~\cite{W28}. 
Subsequently, the idea that CRs escaping from their accelerators follow magnetic 
field lines and preferentially light nearby surrounding molecular clouds 
aligned with them has also been discussed in Ref.~\cite{Malkov:2012qd}. 
Ref.~\cite{Nava:2012ga} addressed this question by assuming 
strongly anisotropic diffusion around sources in one direction. 
Recently, \cite{Kistler:2012ag} confirmed our findings~\cite{Giacinti:2012ar} 
and discussed the implications for local $e^{\pm}$ sources.

In this work, we extend our earlier study mainly
in two respects: First, we present results for the propagation of CRs
with energies as low as 3\,TeV. This allows us to calculate the photon 
images of CR sources down to $E_{\gamma} \geq 300$\,GeV, which is well inside
the energy range accessible to atmospheric Cherenkov telescopes. 
Second, we perform simulations both for protons and electrons,
and discuss differences in the resulting photon images.

After release from their accelerators, CRs initially follow the local magnetic flux tubes, see also~\cite{Malkov:2012qd}, and their density displays filamentary structures. The density distributions still remain irregular and anisotropic on large distances from the sources---up to ${\cal O}(l_{\max})$, even when no collimated filament is visible. In some cases, this may lead to wrong identifications of sources. Our findings call for a revision of the standard assumption of isotropic diffusion used in studies of gamma-ray emission from molecular clouds surrounding sources. We also propose to use gamma-ray observatories as new probes of the interstellar turbulent magnetic fields. We find that gamma-ray spectra for emissions from filaments are steeper than those from surrounding regions, both for CR protons and electrons---provided that CRs in the considered energy range have escaped from the accelerator.

This article is structured as follows. Section~\ref{Code} contains a description of our numerical methods, a list of simplifying hypotheses and their justifications. Sections~\ref{LocalMF}, \ref{AnisoDiff} and \ref{GRSpectra} present the results of this study: We discuss in Sec.~\ref{LocalMF} how the structure of the local turbulent magnetic field is reflected in the gamma-ray images of CR sources. We show in Sec.~\ref{AnisoDiff} simulations of extended gamma-ray emissions, and discuss how the CR radial distributions around sources depart from the predictions of isotropic diffusion. In Sec.~\ref{GRSpectra} we compute the energy spectra of gamma-rays emitted by CR protons and electrons. Finally, we conclude and discuss perspectives in Sec.~\ref{Conclusions}.

\section{Description of the code and assumptions}
\label{Code}

We propagate individual CRs in turbulent magnetic fields ${\vec B}({\vec k})
\propto \exp(-{\rm i}\vec k\vec x)$ using the numerical code described in 
Refs.~\cite{nuc2,nuc3}. The main ingredient for the calculations of
the trajectories of charged particles is the spectrum $\mathcal{P}(\vec k)$ 
of magnetic field fluctuations which is currently only poorly known. 
In principle, it may deviate from purely isotropic turbulence, as suggested
by theoretical models as, for example, Ref.~\cite{Goldreich:1994zz}. In this
work we will use an isotropic spectrum of fluctuations because the main 
effect---strongly anisotropic diffusion on scales $\lesssim 
{\cal O}(l_{\max})$---will not be weakened by an anisotropy in 
$\mathcal{P}(\vec k)$. Moreover, we assume the power-spectrum to be static
and to follow a power-law, $\mathcal{P}(k)\propto k^{-\alpha}$. 
Although using static fields would not be suitable for some specific studies, see~\cite{O'Sullivan:2009sc,Beresnyak:2013ria}, this assumption is well justified in our case. Both the Alfv\'en velocity $v_{\rm A}$ and the velocity of the interstellar medium $u$ are $\sim$ tens of km/s, which only induces negligible changes within $\sim (100\,{\rm pc})^{3}$ on the time scales we consider, $t\sim t_{10k}=10$\,kyr: The length $t_{10k}\times\max(v_{\rm A},u)$ is at least one order of magnitude smaller than the width of the thinnest CR filaments. 
We use for the normalization of the root mean square (rms) magnetic field strength $B_{\rm rms}^2\equiv \langle {\vec B}^2({\vec r})\rangle=4\,\mu$G, and a fluctuation
spectrum bounded by $l_{\min}=1$\,AU and $l_{\max}=150$\,pc. In the numerical simulations, we use nested grids as 
described in~\cite{nuc3}. This allows us to choose an effective $l'_{\min}=5 \cdot 10^{-4}$\,pc~\cite{nuc3} sufficiently 
small compared to the Larmor radius $r_{\rm L}=cp/(eB)$ of CRs with energies down to $cp\sim 3$\,TeV.

\begin{figure}
\begin{center}
\includegraphics[width=0.49\textwidth]{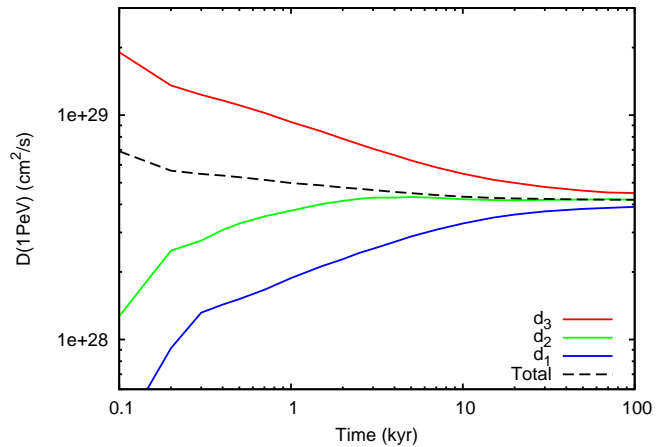}
\end{center}
\caption{Eigenvalues $d_i$ (solid lines) of the diffusion tensor $D_{ij}=\langle x_ix_j\rangle/(2t)$. Black dashed line for the average diffusion coefficient $D$ as a function of time $t$. For a Kraichnan spectrum, $B_{\rm rms}=4\,\mu$G, $l_{\max}=150$\,pc and CR energy $E=10^{15}$\,eV.}
\label{EigenvaluesKr}
\end{figure}

The spectral index $\alpha$ of the turbulent GMF is only 
weakly constrained, and both Kolmogorov ($\alpha = 5/3$) and Kraichnan 
($\alpha = 3/2$) spectra are consistent with observations~\cite{galprop,obs}. 
We compute in Fig.~\ref{EigenvaluesKr} the eigenvalues of the diffusion tensor 
$D_{ij}=\langle x_ix_j\rangle/(2t)$ as a function of time $t$, for 1\,PeV CRs 
in Kraichnan turbulence, averaged over 10 magnetic field configurations. By comparing 
it with Fig.~1 from~\cite{Giacinti:2012ar} 
which was calculated for Kolmogorov turbulence, one can see that results do not vary much 
between the two cases. Therefore, we present most of our remaining results
for the case of a Kolmogorov spectrum. Unless otherwise stated,
we will always use in the following numerical examples 
Kolmogorov turbulence with $l_{\max}=150$\,pc, $l_{\rm c}=l_{\max}/5=30$\,pc,
and $B_{\rm rms}=4\,\mu$G.

In this work, we focus on the gamma-ray emissions around sources that are produced by CRs which have already escaped from them. CRs with $\sim 100$\,TeV energies or higher are believed to be accelerated during the first few hundred years after a SN explosion. Supernova remnants in the Sedov stage may still accelerate CRs up to $\sim (1-10)$\,TeV energies. There is no global consensus on exactly how and when CRs with different rigidities escape from their accelerators---though see~\cite{Bell:2013kq} for a recent study. Since this is not the main topic of the present work, we assume for the sake of simplicity that CRs are released instantaneously at $t=0$. Our results can then easily be convolved with different templates for the rigidity-dependent escape of CRs. For instance, the gamma-ray fluxes due to $\sim 3$\,TeV CRs may then be suppressed at the earliest times and shifted to later times.

Streaming CRs strongly modify the magnetic field in a thin layer in front of a SNR shock, see for example~\cite{Bell:2013kq,Reville:2013bm,Schure:2012du}. The extent to which CRs that have escaped from their accelerator may modify fields within $\sim l_{\rm c}$ from it depends on several unknown parameters, such as the amount of CRs that have effectively been released. In the following computations, we neglect the CR back-reaction on the surrounding interstellar magnetic fields. Even in the 'worst case' scenario, one should not expect significant deviations from our test particle approach for CRs with $E \gtrsim$\,a few tens of TeV ($E_{\gamma} \gtrsim$\,a few TeVs), see Appendix~\ref{App} for a more detailed and quantitative discussion.

We use the continuous energy loss (CEL) approximation when calculating the 
trajectories of CR electrons. Except for sources close to the Galactic
center, the energy losses are dominated by inverse Compton scattering 
(ICS) on CMB photons, because ICS on star-light takes place in the 
Klein-Nishima regime at the energies of interest. Additionally, 
synchrotron losses are important in regions of strong magnetic fields. 
The combined energy losses $dE/dt=\beta E^2$ for an electron are
\begin{equation} \label{CEL}
 \frac{dE}{dt} = 2.5 \times 10^{-18} \,\frac{\rm GeV}{\rm s} 
\left[ \left( \frac{B}{\mu\rm G}\right)^{2} + 9.5 \right] 
\left( \frac{E}{\rm GeV} \right)^{2} \,,
\end{equation}
while the energy losses of protons are negligible on the considered time
scales.

\begin{figure}
\begin{center}
\includegraphics[width=0.49\textwidth]{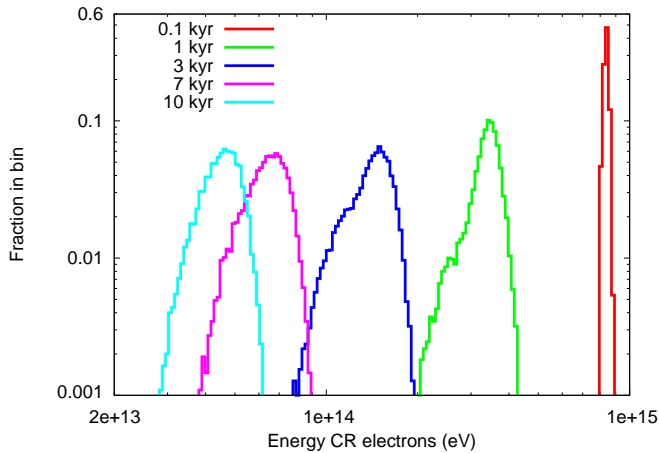}
\end{center}
\caption{Distributions of energies at times $t=0.1,\,1,\,3,\,7,\,10$\,kyr, for CR electrons with initial energy $E=1$\,PeV, and propagated in Kolmogorov turbulence with $l_{\max}=150$\,pc and $B_{\rm rms}=4\,\mu$G. See text for details.}
\label{ElecEDistrib}
\end{figure}

Figure~\ref{ElecEDistrib} shows the evolution with time of CR electron 
energies after being injected at $t=0$ into Kolmogorov turbulence with 
$l_{\max}=150$\,pc and $B_{\rm rms}=4\,\mu$G. Their initial energy 
is $E=1$\,PeV. Although we use the CEL approximation, the electron 
energy distributions widen with time, because the CR 
trajectories probe regions with different magnetic field strengths. Pulsars are 
known to accelerate electrons to PeV energies. On the contrary, electrons 
are not expected to reach such high energies in SNRs, 
because of synchrotron losses during acceleration. For completeness, we show in Section~\ref{GRSpectra} gamma-ray spectra for electrons 
with energies up to 1\,PeV. The cutoff would be shifted towards lower values 
in the case of SNRs.

Cosmic ray electrons produce high-energy gamma-rays mainly by ICS. 
While we employ the CEL approximation to calculate the electron 
trajectories, we derive the photon spectra using the full ICS cross section
and rates extracted from~\cite{elmag}. For an observer located at a 
distance $D$, 
the differential photon flux from IC reactions of CR electrons is given by
\begin{align}
\lefteqn{
 j(E_\gamma,\theta,\phi) = \frac{c}{D^2} \int_{E_\gamma}^\infty dE_e
  R_{\rm ICS}(E_e) }
\nn
\\ & \times
 \frac{1}{\sigma_{\rm ICS}} \,\frac{d\sigma_{\rm ICS}(E_e,E_\gamma)}{dE_\gamma} \,
 \frac{d^3N_e(E_e,\theta,\phi)}{dE_e \,d\Omega} ,
\end{align}
where we assume that the produced photons and the initial electrons move 
collinearly.

The differential number $d^3N_i(E,\theta,\phi)/(dE\,d\Omega)$ of protons and 
electrons with energy $E$ and momentum vector pointing towards the 
observer is normalized to the total energy in CRs, which we choose here 
as $E_{\rm CR}=10^{50}$\,erg. In our examples, we normalize the relative contributions of electrons and protons to the total 
CR flux by requiring that the electron flux is suppressed by a factor 
$K_{\rm ep}\sim 0.3\%$ relative to the proton flux for $E \gg m_p$. 
These values are appropriate for SNRs, while the gamma-ray emission from 
pulsars is dominated by electrons. However, different values of $E_{\rm CR}$ or 
$K_{\rm ep}$ would only change the normalizations of the spectra shown in 
Section~\ref{GRSpectra}.

Cosmic ray protons scattering on the interstellar gas produce secondaries 
which in turn decay into photons. 
An observer at a distance $D$ receives the differential photon flux
\be
 j(E_\gamma,\theta,\phi) = \frac{cn_{\rm H}}{D^2} \int_{E_\gamma}^\infty dE_p
 \frac{d\sigma_\gamma(E_p,E_\gamma)}{dE_\gamma} \,
 \frac{d^3N_p(E_p,\theta,\phi)}{dE_p \,d\Omega} ,
\ee
where we use again that the secondaries move collinearly to the CR primary, 
and $n_{\rm H}$ is the number density of gas. 
The production cross section of photons in p-gas collisions 
has been tabulated by using QGSJET-II-04~\cite{Ostapchenko:2010vb}. 
The increased cross section of heavier nuclei both in the CR 
primary flux and the target gas is accounted for by using a nuclear 
enhancement factor $\eps_m \sim 1.8$.
For the sake of clarity, we assume a uniform gas density $n_{\rm H}$ around sources. We have seen in Ref.~\cite{Giacinti:2012ar} 
that CR momenta in filaments are sufficiently isotropized to assume that the gamma-ray 
surface brightness is proportional to the CR column density. In the following, we use this approximation.

\section{Gamma-ray astronomy as a probe of interstellar magnetic fields}
\label{LocalMF}

\begin{figure*}[!t]
  \centerline{\includegraphics[width=0.49\textwidth]{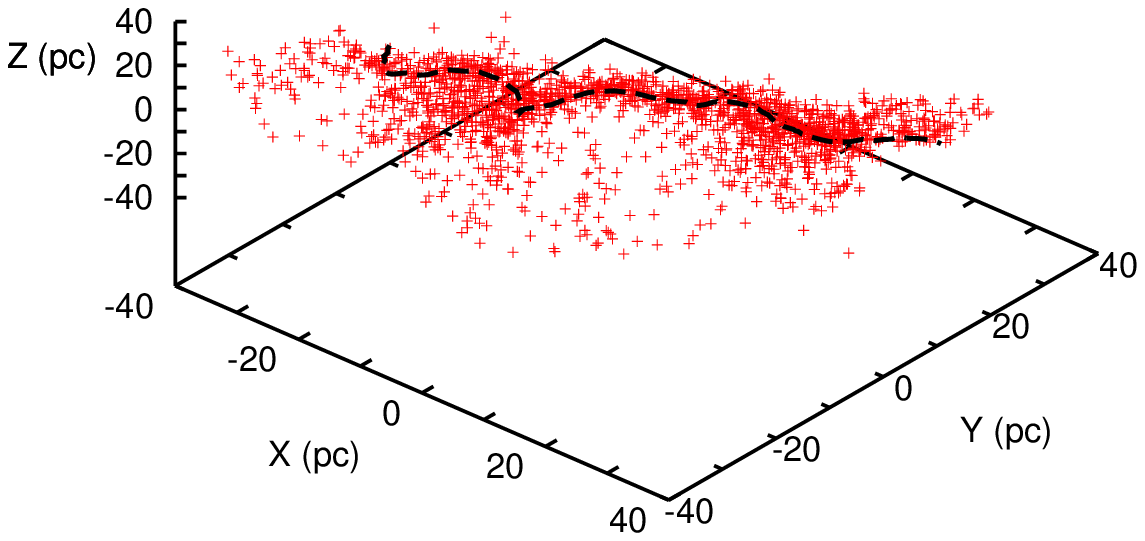}
              \hfil
              \includegraphics[width=0.49\textwidth]{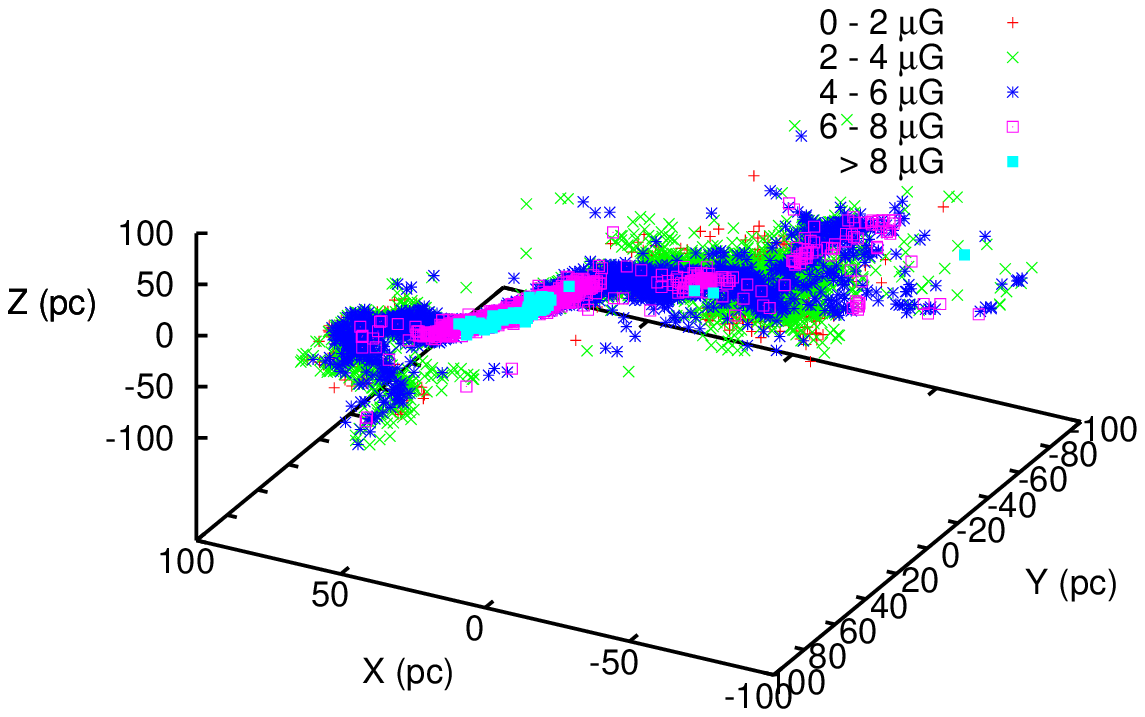}
             }
  \caption{{\it Left panel :} 3D spatial distribution of 1\,PeV protons (red crosses) emitted at (0,0,0) 500\,yr ago. The dashed black line represents the magnetic field line of the local random magnetic field that contains (0,0,0); {\it Right panel :} 3D spatial distribution of 1\,PeV protons emitted at (0,0,0) 1\,kyr ago. Colours of symbols correspond to the local magnetic field strengths at the cosmic ray positions, see key.}
  \label{FieldLines}
\end{figure*}

Assuming $l_{\max} \sim 100$\,pc, Larmor radii of TeV--PeV CRs are at least two orders of magnitude 
smaller than the largest fluctuation scales in the random magnetic field, 
where most of the power resides, $\mathcal{P}(2\pi/r_{\rm L}) \ll \mathcal{P}(2\pi/l_{\max})$. 
Large scale modes with $2\pi/k \gg r_{\rm L}$ act as local regular fields. 
This results in anisotropic diffusion of CRs at early times~\cite{Giacinti:2012ar}. The degree of anisotropy 
depends on the local ratio of power in modes ${\cal O}(\gg r_{\rm L})$ and ${\cal O}(r_{\rm L})$. 
Since the local directions of modes with $2\pi/k \gg r_{\rm L}$ change 
on scales smaller than $l_{\max}$, the shapes of CR distributions look irregular on distances $\lesssim {\cal O}(l_{\max})$, 
see Figs.~\ref{Filaments} and~\ref{EmissionOnTheSky}.

The presence of a regular field ${\vec B}_{\rm reg}$ coherent on 
$\sim$\,kpc scales would not change our results qualitatively. The regular 
field ${\vec B}_{\rm reg}$ simply adds up to the large scale modes. Since
it is weaker than the turbulent field which contains 
most power on large scales, the resulting change of the large scale modes 
is small. The standard description of diffusion in the presence of a uniform magnetic
field, which uses a parallel and a perpendicular diffusion coefficient along its direction, 
only becomes applicable when the bulk of CRs reach 
distances $\gtrsim {\cal O}(l_{\max})$ from their source. We have also checked numerically that results with a regular field are not much different from without, as long as $|{\vec B}_{\rm reg}| \lesssim B_{\rm rms}/2$, as expected in the Galaxy~\cite{Jansson:2012pc,Jansson:2012rt}. For instance, for 1\,PeV protons in $|{\vec B}_{\rm reg}| \sim 2\,\mu$G and $B_{\rm rms} \sim 6\,\mu$G, the presence of a regular field produces at large distances from the source a flux similar to what is observed without it. The only difference is the elongation of the CR distribution along the direction of the regular field at $r>l_{\max}$.

The CR densities around recent sources reflect the structures of the surrounding magnetic fields. Therefore, we propose that gamma-ray astronomy can be used in the future as a new way to probe the still poorly known turbulent magnetic field that permeates the interstellar medium. This field is currently mainly studied through unpolarized synchrotron emission. Gamma-ray astronomy offers a different and complementary perspective. Contrary to synchrotron emission observations, it would be a local probe, within ${\cal O}(l_{\max})$ from CR proton and electron sources. Studies of anisotropic extended gamma-ray emissions around sources would notably allow one to estimate independently important parameters of the turbulence such as $l_{\rm c}$ and $l_{\max}$.

Observations suggest that $l_{\rm c} \sim$ a few tens of parsecs, and $l_{\max} \sim 100$\,pc in our Galaxy~\cite{Han:2004aa}. Around the Solar system, observations of polarized starlight from nearby stars also show that the turbulent field is coherent on scales of at least several tens of parsecs~\cite{Frisch:2012zj,Frisch:2011qf,Frisch:2010fw}. From a theoretical perspective, such values are consistent with those expected if this field is the result of the fluctuation dynamo~\cite{Subramanian:2008tt}. Similar or larger correlation scales are expected in the turbulent flow of the interstellar medium, driven by supernova explosions. See, for instance, Ref.~\cite{Gent:2012eh} for a recent study and~\cite{Armstrong:1995zc} for observational evidence.

Our main results can be summarized as follows:

For short times after escape from their accelerators, typically $t \lesssim 1$\,kyr for $B_{\rm rms }=4\,\mu{\rm G}$, $l_{\max}=150$\,pc and $E/Z=1$\,PeV rigidity, high energy CRs diffuse within a magnetic flux tube containing the source. This results in a filamentary distribution of CRs. Figure~\ref{FieldLines} {\it (left panel)} shows in red the three-dimensional distribution of 1\,PeV protons at $t=500$\,yr. Cosmic rays clearly follow the black dashed line which shows the magnetic field line that goes through the source. As time goes by, the length and, to a lesser extent, the width of the filament increase. The latter effect is caused by magnetic field line wandering. The degree of anisotropy of the CR distribution depends on the local strength of the turbulent magnetic field around the source: Regions of strong magnetic fields correspond to regions where the amplitude of the large scale modes is large. As expected, CR distributions around sources are found to be statistically more filamentary in this case because the ratio of power in the large scales $\gg r_{\rm L}$ to smaller scales $\sim r_{\rm L}$ is larger. On the contrary, CRs are found to spread less anisotropically around sources that are located in regions with weaker fields. The same effect can also result in a higher or lower degree of collimation of a specific filament along its axis. We plot in Fig.~\ref{FieldLines} {\it (right panel)} the spatial distribution of 1\,PeV protons 1\,kyr after their emission from a source located in (0,0,0). The colour of each CR corresponds to the strength of the field it experiences, see the key in the figure. The thinnest and most collimated section of the filament is situated in the region with the strongest fields. Thus, gamma-ray observations of filaments can provide in principle information on the local field strength, as well as on the geometry of field lines.

Statistically, one would not expect very filamentary gamma-ray emissions on lengths longer than about ${\cal O}(l_{\rm c})$ from the sources. Directions of field lines that are within a few parsecs from one another around the centers of sources do not remain correlated on much larger scales. This results in a larger spatial spread of escaping CRs. The radii of historical supernova remnants or remnants in the early Sedov phase do not exceed a few parsecs, which should be $\ll l_{\rm c}$. Therefore, filamentary diffusion of CRs should be statistically expected around at least some of them, except if they are located in atypical regions with $l_{\rm c},\,l_{\max} \lesssim $ a few\,pc. For sources whose sizes are comparable with or larger than $l_{\rm c}$, such as remnants in the late Sedov phase, the gamma-ray emissions around them that result from escaping CRs are still expected to be anisotropic and to follow field lines, though the simple picture of CRs confined in one flux tube does not hold any more.

At later times after CR escape, typically $1\,{\rm kyr} \lesssim t \lesssim t_{\ast} = 10$\,kyr for $E/Z=1$\,PeV, CRs start to explore regions beyond $r \sim l_{\rm c}$ from the source. Even if their distribution in space is less anisotropic than at early times, a non-negligible departure from the predictions of standard isotropic diffusion still remains until they reach distances $r$ from the source equal to about one or two $l_{\max}$. The resulting extended gamma-ray emission is expected to be anisotropic typically until 
\be \label{tast}
t_{\ast} \sim 10^4 \,{\rm yr}\; \left( l_{\max}/150\,{\rm pc}\right)^{\beta}
\left( E/{\rm PeV} \right)^{-\gamma}
\left(  B_{\rm rms} /4\,{\rm \mu G} \right)^{\gamma}
\ee
with $\beta \simeq 2$ and $\gamma=0.25$--$0.5$ for Kolmogorov turbulence, 
see~\cite{Giacinti:2012ar}. During this period, anisotropic diffusion still
leads to important consequences for gamma-ray astronomy which will
be discussed in the next Section. 
The transition time $t_{\ast}$ corresponds to the time needed for CRs to randomize in a $\sim$ few $l_{\max}^3$ volume around the accelerator, and reach the distribution expected from isotropic diffusion predictions. Even though one cannot realistically retrieve information on the magnetic field lines any more, precious information on the generic properties of the random magnetic field such as $l_{\max}$ can still be deduced using Eq.~(\ref{tast}).
Therefore, studying the degree of anisotropy of extended gamma-ray emissions around sources, depending on their age and location in the Galaxy, will yield, in the future, information about characteristic parameters of the interstellar turbulent magnetic fields such as $l_{\max}$ and $B_{\rm rms}$. In principle, it could also give insights into their power spectrum.

As stated in~\cite{Giacinti:2012ar}, flatter turbulence spectra such as a Bohm spectrum result in less anisotropic CR diffusion. This is due to the fact that more power is present in modes with variation scales ${\cal O}(r_{\rm L})$. For a given $l_{\max}$, $l_{\rm c}$ is smaller than for a Kolmogorov spectrum, and then nearby field lines remain coherent on smaller scales. This results in a faster randomization of CR positions in space.

\section{Extended gamma-ray emissions around sources and departure from isotropic CR diffusion}
\label{AnisoDiff}

\begin{figure*}[!t]
  \centerline{\includegraphics[width=0.33\textwidth]{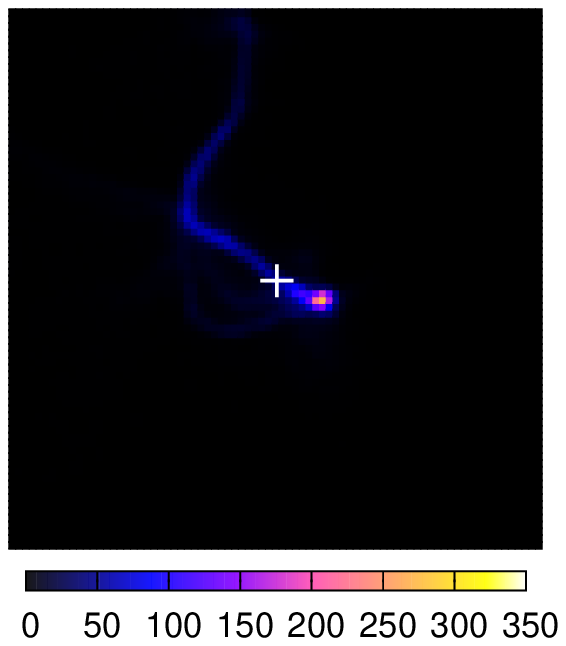}
              \hfil
              \includegraphics[width=0.33\textwidth]{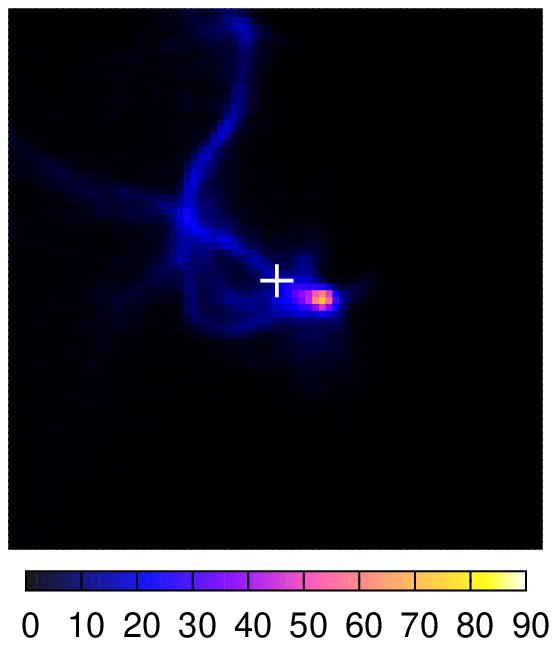}
              \hfil
              \includegraphics[width=0.33\textwidth]{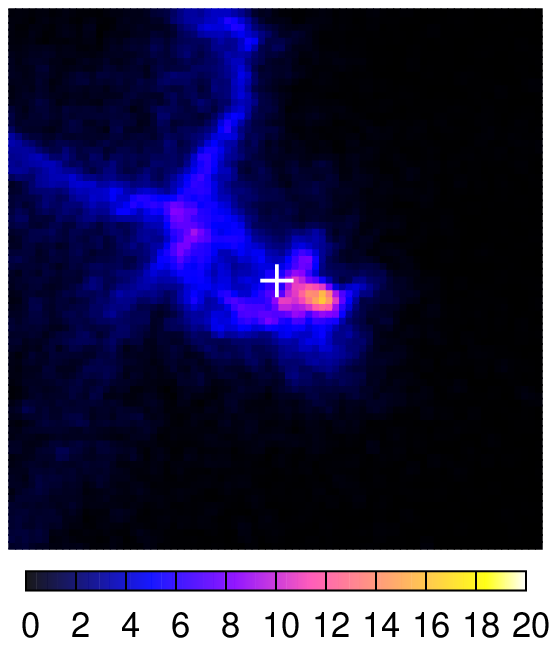}
             }
  \caption{Relative surface brightness in gamma-rays on the sky above $E_{\gamma} \geq 300$\,GeV for a proton source --See key below each panel. $t=1$\,kyr ({\it left panel}), $t=3$\,kyr ({\it middle panel}) and $t=7$\,kyr ({\it right panel}) after escape from the source. An $E^{-2}$ spectrum and a maximum energy $E_{\max}=1$\,PeV have been assumed for the source. For clarity, the density of target gas is supposed to be homogeneous. Same parameters are used as in previous figures for the interstellar magnetic fields. Each panel size corresponds to an 80\,pc$\times$80\,pc region. The white cross represents the position of the source.}
  \label{Filaments}
\end{figure*}

\begin{figure*}[!t]
  \centerline{\includegraphics[width=0.33\textwidth]{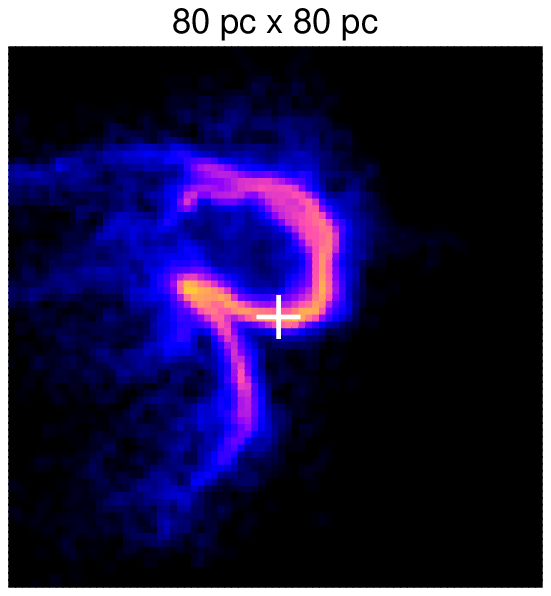}
              \hfil
              \includegraphics[width=0.33\textwidth]{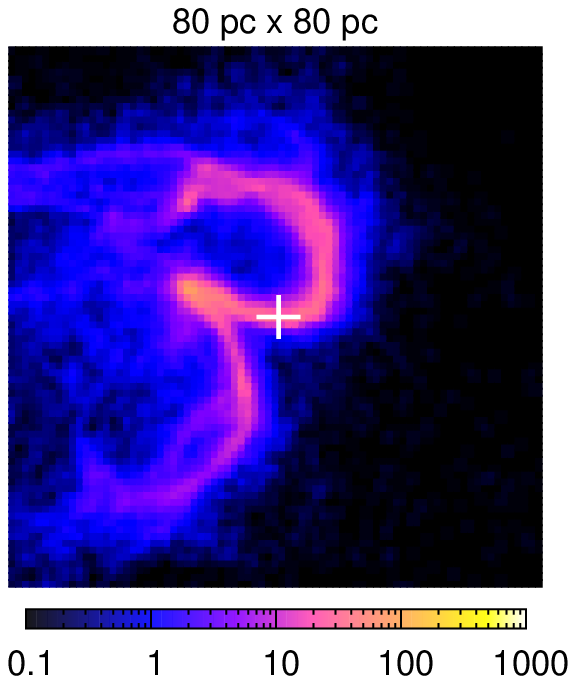}
              \hfil
              \includegraphics[width=0.33\textwidth]{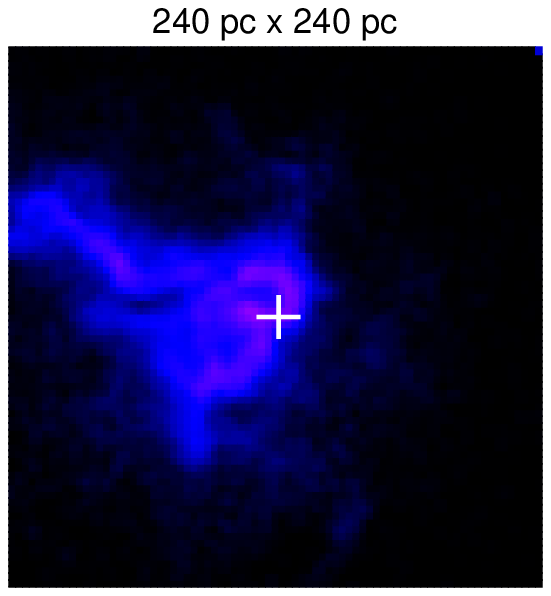}
             }
  \centerline{\includegraphics[width=0.33\textwidth]{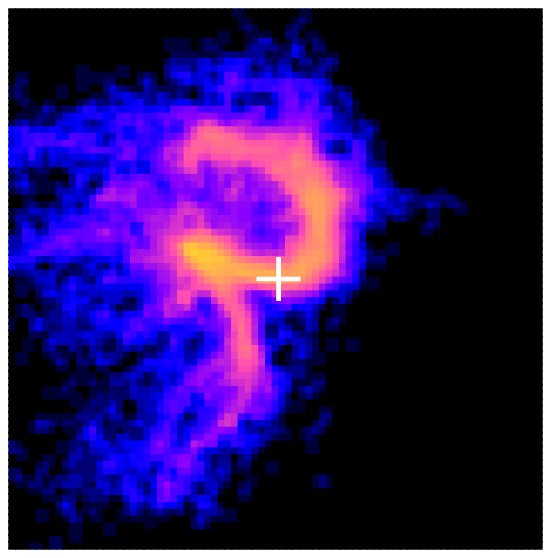}
              \hfil
              \includegraphics[width=0.33\textwidth]{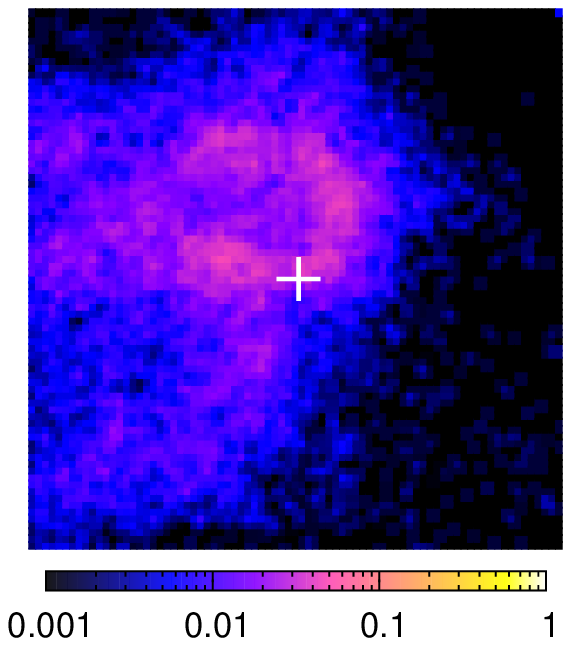}
              \hfil
              \includegraphics[width=0.33\textwidth]{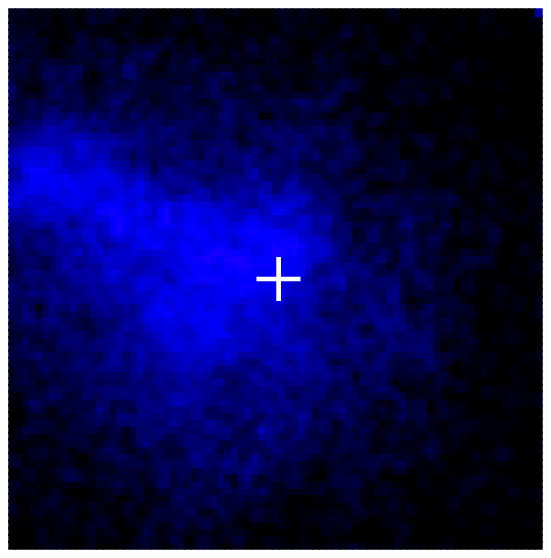}
             }
  \caption{Relative surface brightness in gamma-rays on the sky above $E_{\gamma} \geq 300$\,GeV ({\it upper row}) and $E_{\gamma} \geq 30$\,TeV ({\it lower row}) for a proton source, at $t=0.5,\,2,\,10$\,kyr after escape ({\it resp. left, middle and right columns}). For each row, see key below the middle panel. Same parameters and assumptions are used as for Fig.~\ref{Filaments}. White cross for the position of the source. The panel sizes are 80\,pc$\times$80\,pc ({\it left and middle columns}) and 240\,pc$\times$240\,pc ({\it right column}). Note the significant offset between the source position and the center of the extended gamma-ray emission on the two right panels.}
  \label{EmissionOnTheSky}
\end{figure*}

The authors of Ref.~\cite{Nava:2012ga} modeled the diffusion of CRs close to their
source by an anisotropic diffusion tensor, resulting in faster diffusion along a given axis.
Given the sensitivity and angular resolution of present gamma-ray 
observatories, and the amount of poorly constrained astrophysical parameters, 
such a treatment is justified.  However, the true CR distributions are
significantly more complicated in general, and the better angular resolution 
expected, for instance, for H.E.S.S.~II~\cite{hessii}, MAGIC-II~\cite{magicii} and the CTA observatory~\cite{Consortium:2010bc} will 
reveal more subtle structures in the emission patterns. Filaments are 
rarely expected to appear as clear and straight lines expanding in directly 
opposite sides of the source. In some cases, the center of the extended gamma-ray emission may have a significant offset from the location of the source. This can be seen in Figure~\ref{EmissionOnTheSky} which shows the relative surface brightness in gamma-rays on the sky around a source marked by a white cross. In all panels, most of the gamma-ray emission comes from one side of the source. In practice, this can lead to the misidentification of the CR source, if effects from anisotropic diffusion are neglected when analyzing gamma-ray data.

From one source to another, the degree of anisotropy and the strength of filaments in the CR density vary considerably, depending on the local magnetic field structure. Given their wide variety of shapes, we use the word ``filament'' for any collimated and elongated structure, without making any statement on their degree of collimation.

Figure~\ref{Filaments} shows the relative surface brightness in gamma-rays on the celestial sphere around one given proton source (white cross). Results at three different times after CR escape (1, 3 and 7\,kyr) are presented. We use an $E^{-2}$ spectrum and a maximum energy $E_{\max}=1$\,PeV. We remind the reader that in reality CRs with different energies are expected to escape from the source at different times. Figure~\ref{EmissionOnTheSky} shows the emission at $t=0.5,\,2,\,10$\,kyr after escape (resp. left, middle and right columns). Two different thresholds $E_{\gamma}$ for the gamma-ray energies are considered: $E_{\gamma} = 300$\,GeV for the upper row and $E_{\gamma} = 30$\,TeV for the lower row. The field-of-view is 80\,pc$\times$80\,pc for the two first columns and 240\,pc$\times$240\,pc for the last one.

In the case of Fig.~\ref{Filaments}, one can see several filaments in gamma-rays at $\geq 3$\,kyr times. At 3\,kyr, on the left-hand side of the source, at least three distinct filamentary structures are visible. Also, the maximum emissivity does not correspond to the source position, which is contrary to naive expectations. The filament at 1\,kyr already displays a very bright spot on its right end. It does not fade away and is still present at 7\,kyr. In a realistic situation, the source would still be likely to contain a sufficiently large amount of CRs to outshine the bright blob in gamma-rays on its right. However this example shows clearly that brightness in gamma-rays does not necessarily decrease with the distance to the source, even along one given filament. In other words, even with a homogeneous density of target protons, the gamma-ray emission around one given source can be, in some cases, stronger at larger radii from the source than at smaller ones.

Different factors are responsible for such complicated patterns. Filaments are often significantly twisted, because the local direction of the field varies. Also, two-dimensional projection effects on the sky can make them look significantly different, if not unrecognizable, from naive expectations. For instance, in Fig.~\ref{Filaments} (left panel) the filament is curved towards the observer on its right side, resulting in the bright spot. In Fig.~\ref{Filaments} (middle panel), the apparently multiple filaments arise partly because of foldings of a single filament combined with 2D-projection effects. Only sometimes is the appearance of several filaments not due to projection effects: In a few cases, filaments ``split'' into collimated sub-filaments when field lines in the flux tube containing the CRs 'split' into different sets. Within all tested magnetic field realizations, we found a few ``octopus-like'' sources, displaying several filaments around them.

In Fig.~\ref{EmissionOnTheSky} (upper middle panel notably), the bright pattern shaped as a ``3'' corresponds to the two-dimensional projection of a magnetic flux tube containing the source. One can see that the filament is strongly curved and drives most CRs to one side of the source. Interestingly, at later times (10\,kyr, upper and lower left panels), the gamma-ray emission still mostly comes from the same side of the source, even if CRs are not confined in a thin collimated filament any more. The fact that CRs have been initially driven to one side of the source still strongly influences the shape and location of the emission, as long as CRs are within ${\cal O}(l_{\max})$ distances from the source. From an observational perspective, the extended gamma-ray emissions at $t=10$\,kr (see left column of Fig.~\ref{EmissionOnTheSky}) are not centered on the source position. The source is only at the border. If other potential CR sources are present in the field-of-view, this poses a substantial risk of misidentification of the source. This may explain cases such as that of HESS J1303-631 and PSR 1301-6305~\cite{Aharonian:2005rv,Abramowski:2012fc}.

The two rows in Fig.~\ref{EmissionOnTheSky} show the impact of the gamma-ray energy threshold $E_{\gamma}$ on the emission shape. As expected, for lower $E_{\gamma}$ values, the emission is more strongly filamentary, even if CRs of different energies were to be released at the same time. This is due to the fact that the gamma-ray emissions for $E_{\gamma} \geq 300$\,GeV and $E_{\gamma} \geq 30$\,TeV are dominated, respectively, by that of CRs with $E \sim 3$\,TeV and $E \sim 300$\,TeV. Lower energy CRs diffuse more slowly and therefore stay in the filament for longer times. In practice, the fact that lower energy CRs are expected to escape from the accelerator later than higher energy ones would amplify this effect.

We have computed the same figures for CR electrons instead of protons. With a turbulent field strength $B_{\rm rms}=4\,\mu$G, results for electrons are not significantly different, except at the very highest energies $E_{\gamma} \geq 30$\,TeV. Indeed, energy losses of CR electrons with energies $\gtrsim 100$\,TeV become substantial in a $4\,\mu$G field on kyr time scales, see Fig.~\ref{ElecEDistrib}. For a given local magnetic field realization, the {\em shape\/} of the extended gamma-ray emission from electrons does not significantly differ from that due to protons, except for much stronger fields where it would be spatially less extended than the hadronic one.

\begin{figure*}[!t]
  \centerline{\includegraphics[width=0.235\textwidth,angle=-90.]{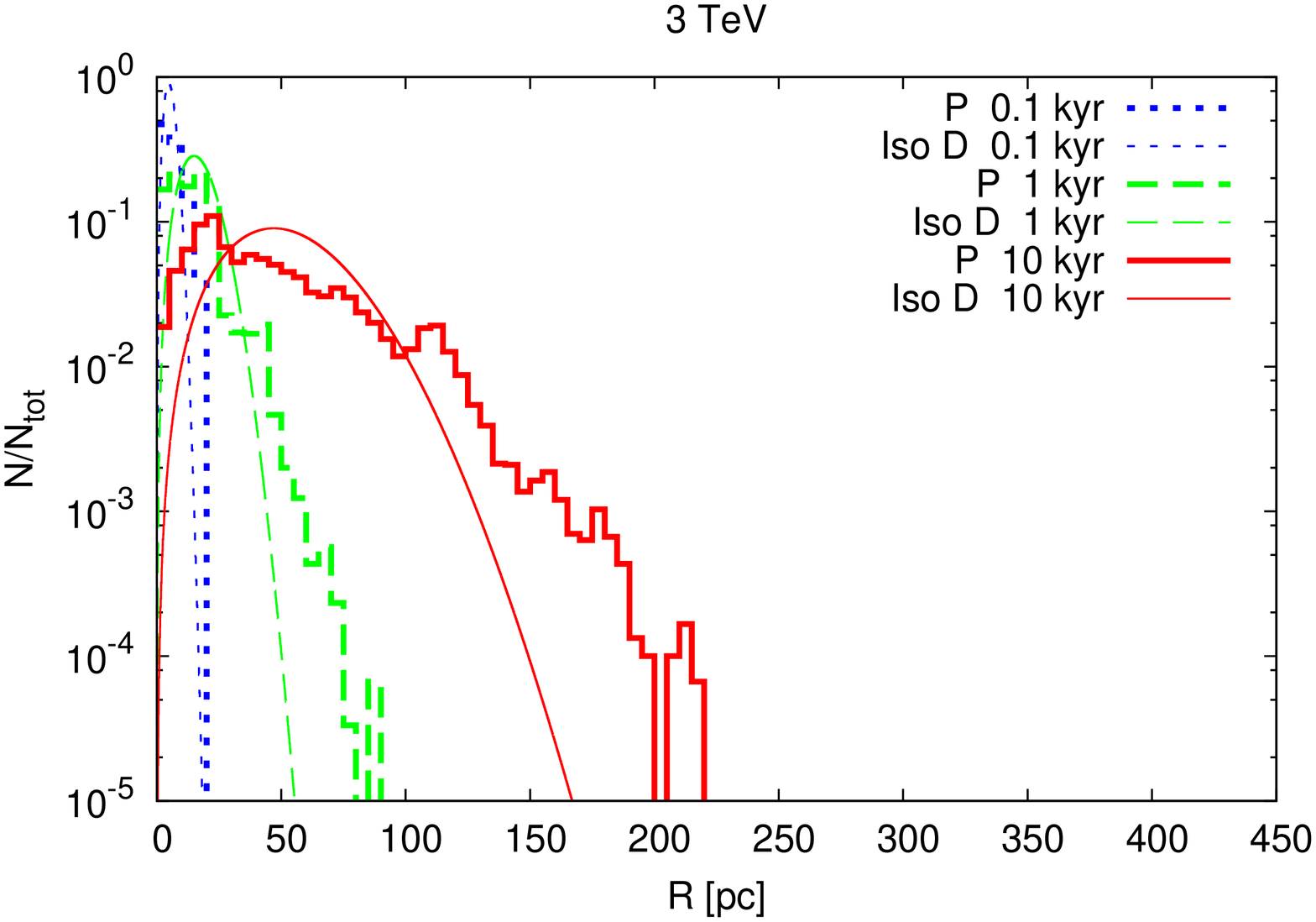}
              \hfil
              \includegraphics[width=0.235\textwidth,angle=-90.]{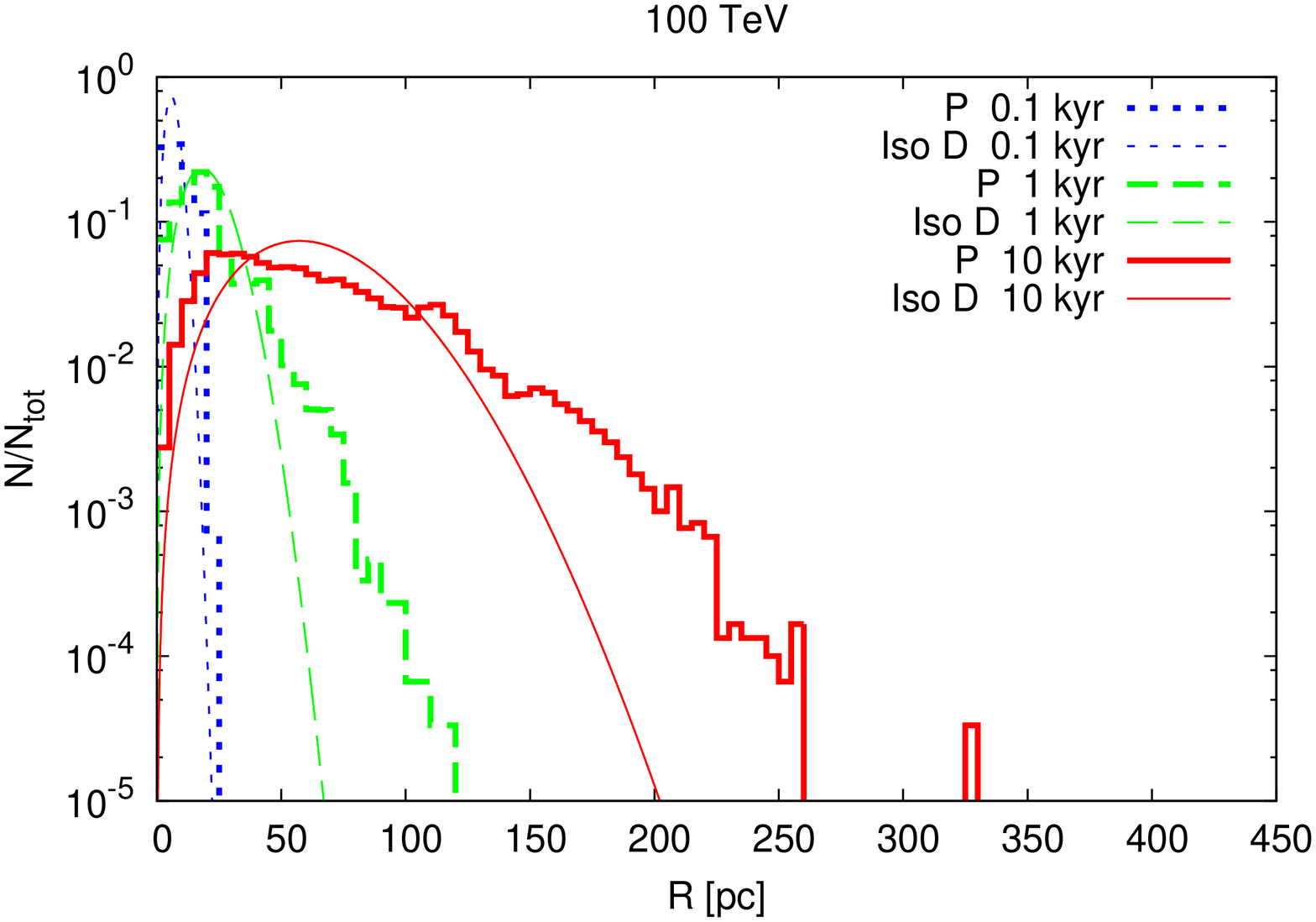}
              \hfil
              \includegraphics[width=0.235\textwidth,angle=-90.]{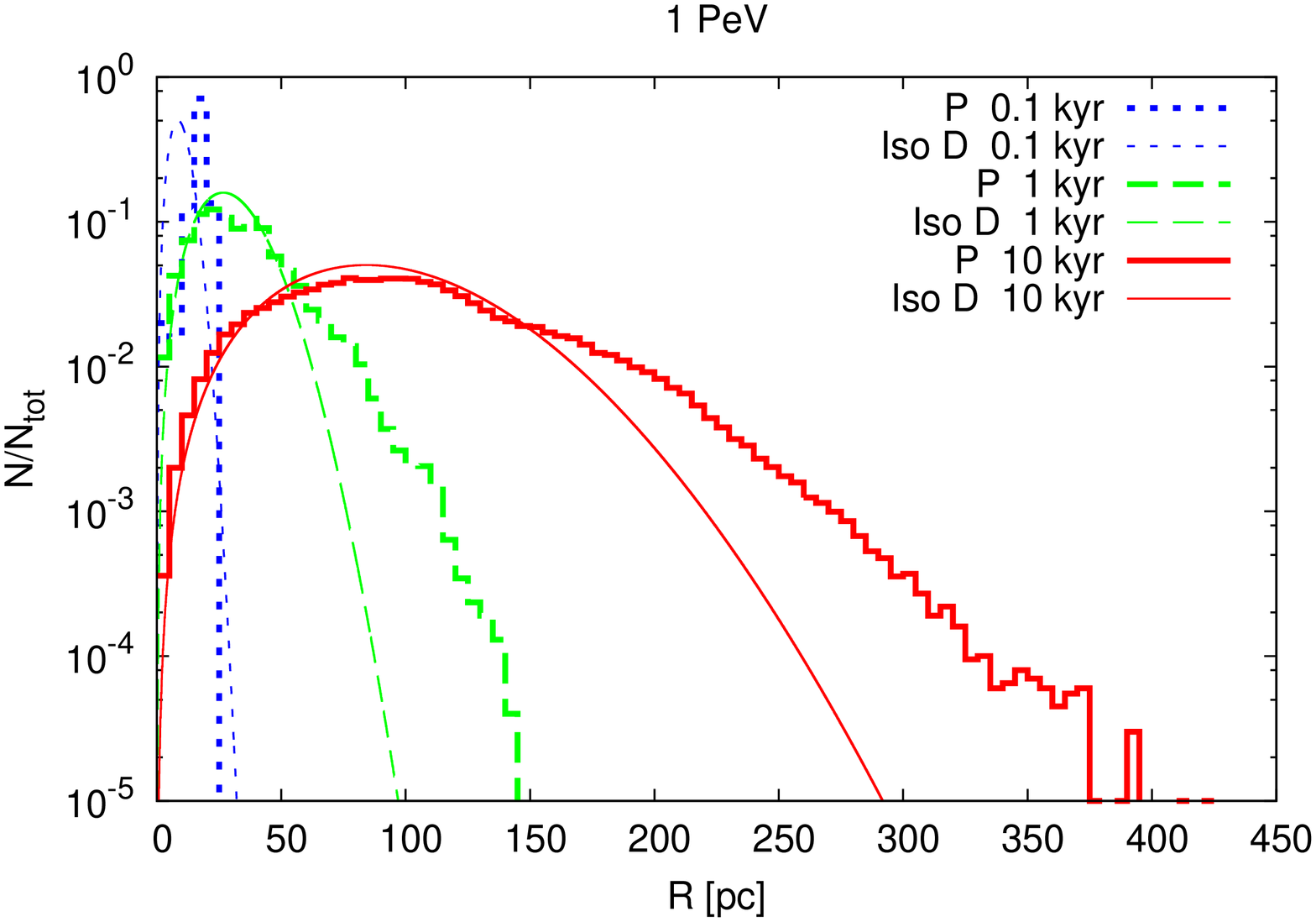}
            }
  \centerline{\includegraphics[width=0.235\textwidth,angle=-90.]{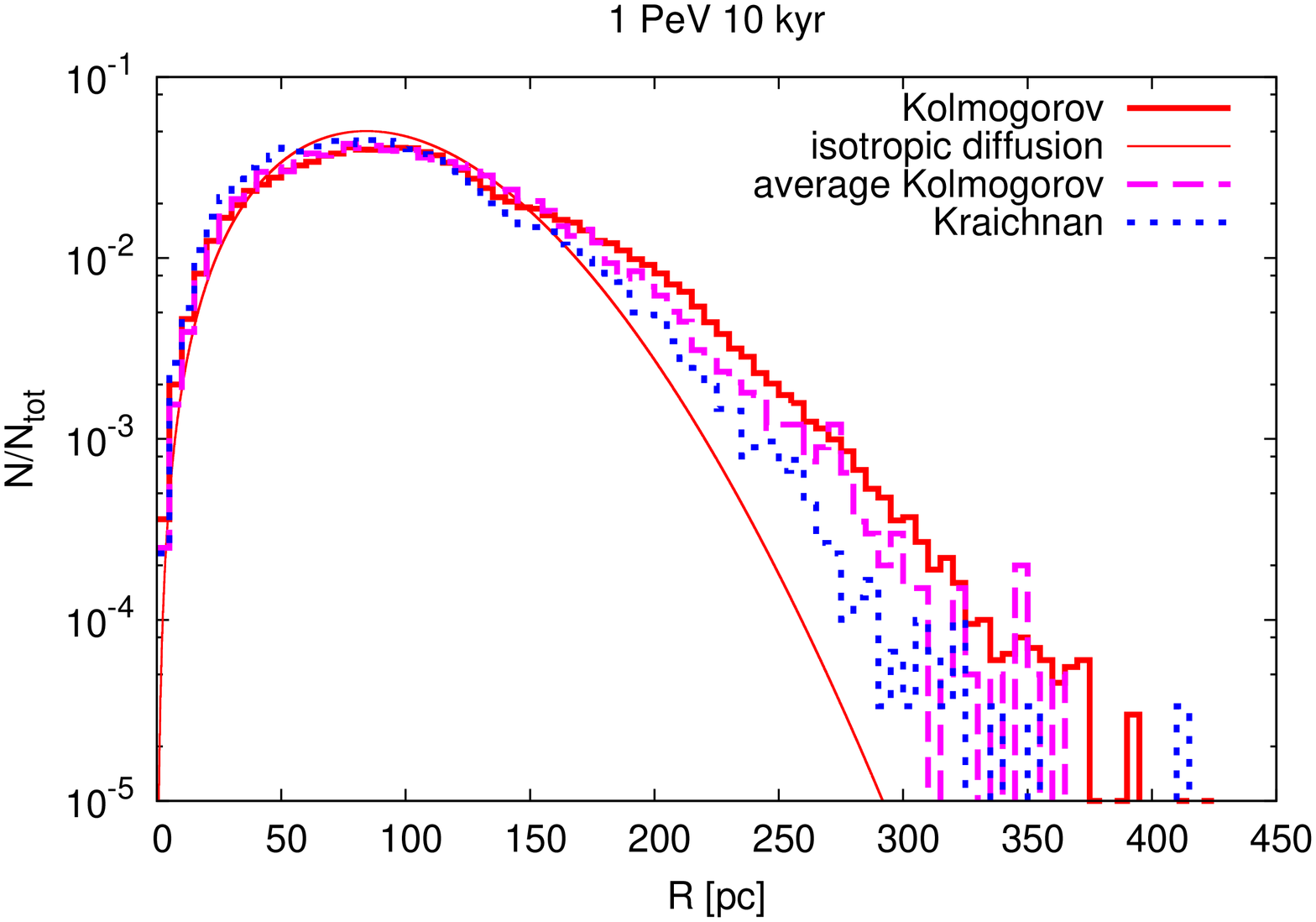}
              \hfil
              \includegraphics[width=0.235\textwidth,angle=-90]{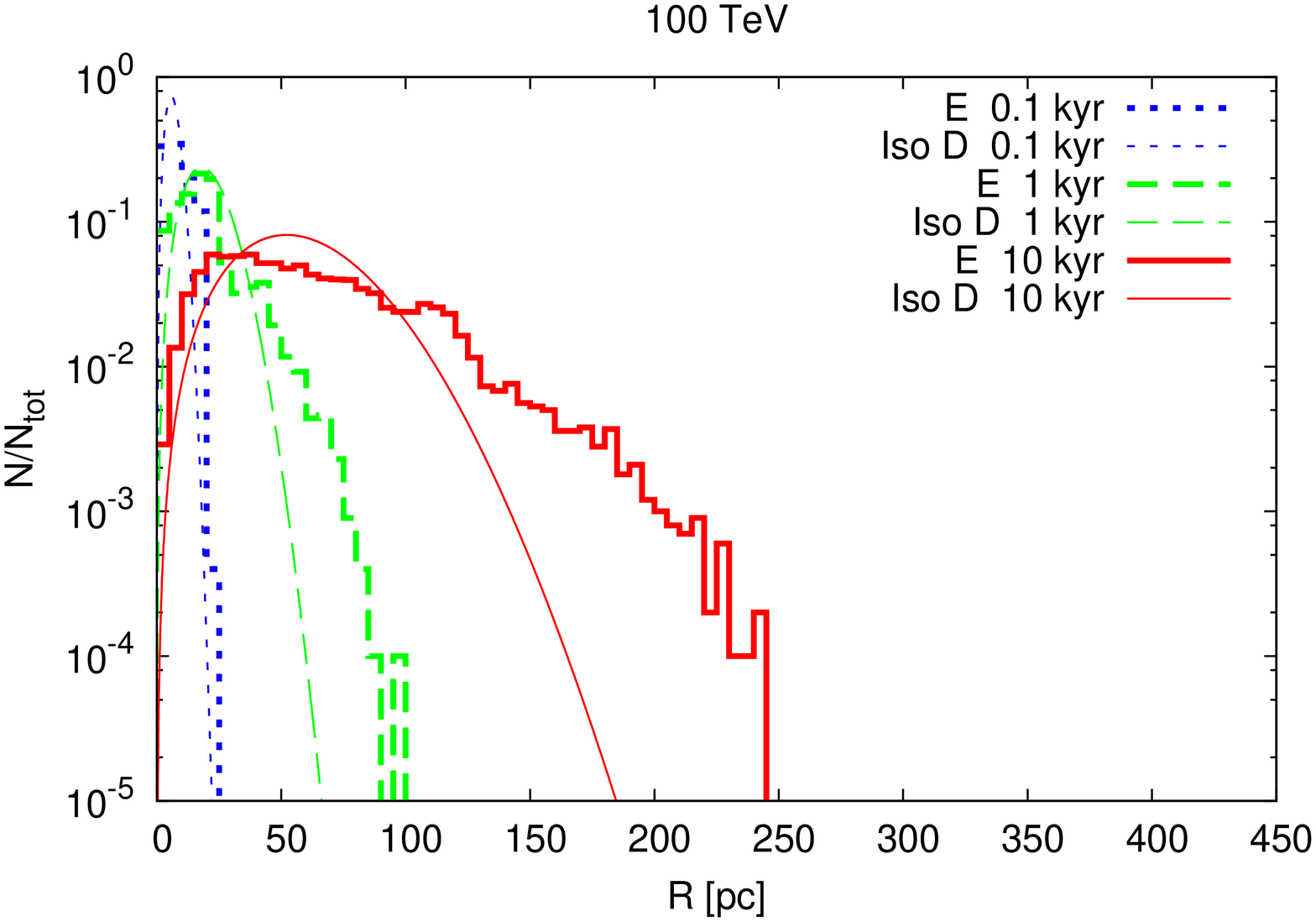}
              \hfil
              \includegraphics[width=0.235\textwidth,angle=-90]{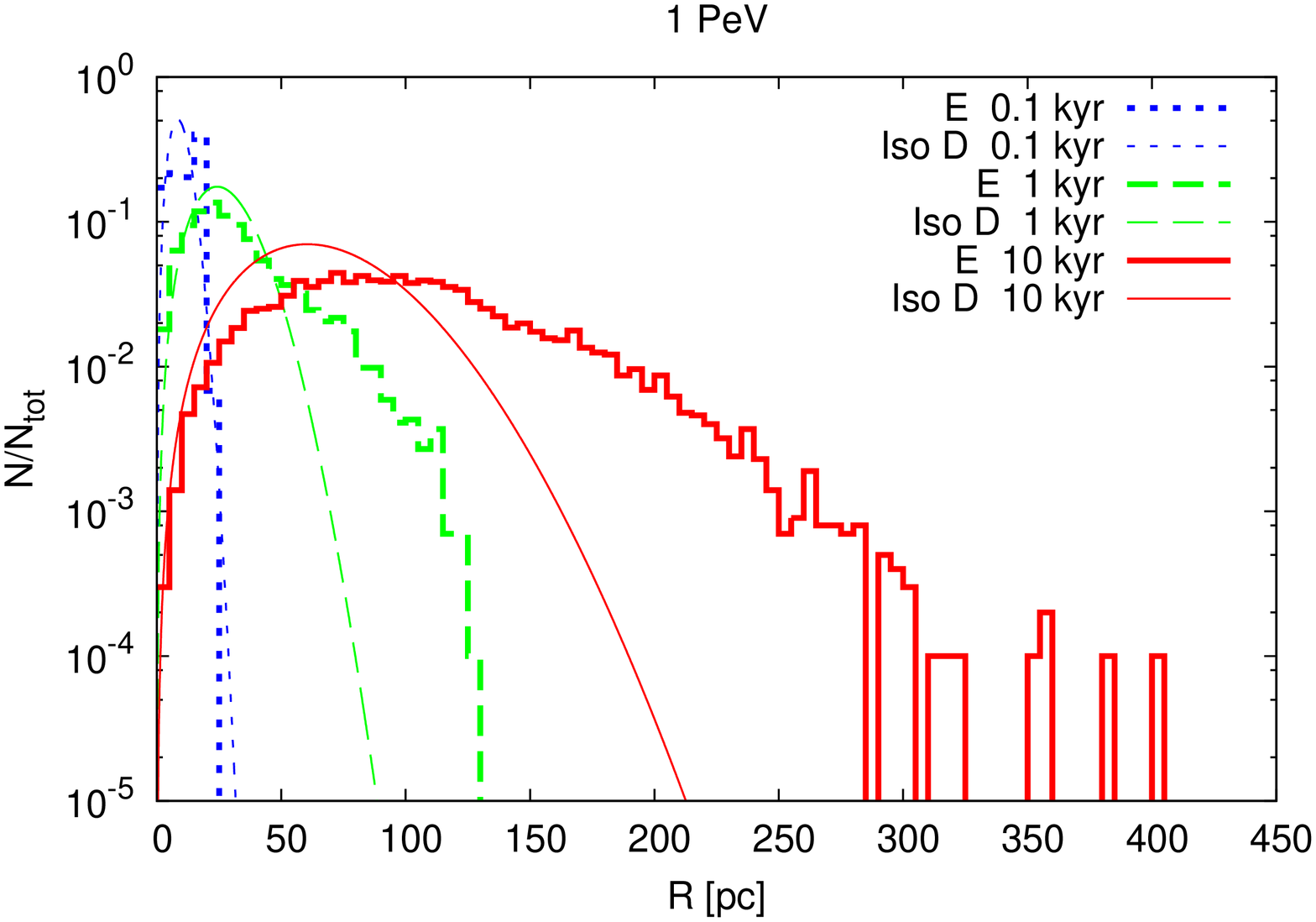}
            }
  \caption{{\it Upper panels:} Radial distributions at $r=0$ and $t=0.1$, 1 and 10\,kyr, of $E=3$, 100, and $1000$\,TeV protons (thick lines) initially injected at $r=0$, compared to isotropic diffusion (thin lines). {\it Lower left panel:} Comparison of the 1\,PeV proton radial distributions at $t=10$\,kyr, for the cases of Kolmogorov turbulence (thick red line), isotropic diffusion (thin red line), Kolmogorov turbulence averaged over 10 different magnetic field configurations (dashed magenta line) and Kraichnan turbulence (dotted blue line). {\it Lower middle and right panels:} Radial distributions of $E=100$ and 1000\,TeV electrons (thick lines) compared to isotropic diffusion (thin lines), assuming same parameters as for the protons.}
  \label{CR_RadialDistr}
\end{figure*}

In Fig.~\ref{CR_RadialDistr}, we show the radial distributions $n(r)$ of CRs 
as a function of the distance $r$ to their source for three times 
$t=0.1$, 1 and 10\,kyr. In the three upper panels, we present the 
distributions $n(r)$ for 3, 100 and 1000\,TeV protons 
(thick lines), injected at $r=0$ and $t=0$ in a turbulent 
magnetic field. Additionally, we show with thin lines the expected 
distributions $n(r)$ of CRs around a bursting source in the isotropic 
diffusion picture~\cite{Ginzburg:1990sk},
\begin{equation}
n(r)\propto \lambda(E,t)^{-3/2}\exp[-r^2/(4\lambda(E,t))] \,
\label{n_3D}
\end{equation}
where $\lambda(E,t)$ is the Syrovatskii variable 
$\lambda(E,t)=\int_0^t dt\,D(E)$. Here we use 
$D(E)= D_0(E/E_0)^{\delta }$ with 
$D_0 =5.5\times 10^{26}{\rm cm/s^2}$, $E_0=1$\,GeV
and $\delta=1/3$ (appropriate value of $\delta$ for Kolmogorov turbulence). This corresponds 
to the diffusion coefficient found in~\cite{Giacinti:2012ar} for such magnetic field parameters, in the limit $t \gg t_\ast$. 
For the electrons, we implement energy losses as specified in~(\ref{CEL}).
The energy of an electron injected at $t=0$ with energy $E_g$ diminishes 
with time as $E(t) = E_g/(1+\beta E_gt)$ and the Syrovatskii variable
evolves as
\begin{equation}
 \lambda^2 =  \left(\frac{E_g}{E_0}\right)^{\delta -1}  
              \left[1-\left(\frac{E}{E_g}\right)^{\delta -1}\right] 
              \frac{D_0}{(\delta -1)\beta(E_g)E_0 } \,.
\end{equation}

For the time scales we are interested in, $t \gg 100$\,yr, the tails of the
true CR distributions shown in Fig.~\ref{CR_RadialDistr} extend to much 
larger distances than expected from the isotropic prediction. In particular, 
the results for $t=10$\,kyr suggest that the true CR distribution 
for large $r$ is not exponentially suppressed but develops a power-law tail. 
For instance, for 3\,TeV protons, one would expect 
only $\simeq 10^{-5}$ of them to be at distances $r>150$\,pc from the source at $t=10$\,kyr, 
while we find this fraction to be $\sim 10^{-3}$, when propagating them in such turbulent fields. 
This is due to the fact that, within a few $l_{\max}$ from the source, the CR distribution still bears 
some imprints of a significantly faster diffusion from the source along filaments at early times, in the first cell of 
size ${\cal O}(l_{\max})$. Moreover, we find that CRs in the tail at large distances from the source are still 
distributed quite anisotropically in space, see for instance the left column of 
Fig.~\ref{EmissionOnTheSky}. As a result, the deviation from the 
isotropic expectation $n(r)$ can reach factors $\sim 10^{4}$--$10^{5}$ at large $r$, in some 
directions around the source.

In the lower left panel of Fig.~\ref{CR_RadialDistr}, we compare the radial distributions $n(r)$ 
of 1\,PeV protons at $t=10$\,kyr, for two different cases of turbulent field spectra, Kolmogorov ($\alpha=5/3$, 
thick red line) and Kraichnan ($\alpha=3/2$, dotted blue line). 
As expected, the tail of $n(r)$ at large distances is reduced for a Kraichnan 
spectrum in comparison with a Kolmogorov one. Nevertheless, even for Kraichnan 
turbulence the tail extends to larger distances than predicted by isotropic 
diffusion. We also show the radial distribution $n(r)$ averaged over ten magnetic field configurations 
with Kolmogorov spectra (dashed magenta line). The agreement with the previous computations demonstrates that the 
magnetic field configuration chosen for the above results with $\alpha=5/3$ is not atypical.

Finally, we present results for electrons with initial energies $E=100$ and 
$1000$\,TeV in the lower middle and right panels. Our 
results (thick lines) are again compared to the predictions of isotropic diffusion 
(thin lines). The tails at large distances $r$ 
are also present for electrons. The deviation from the isotropic expectations 
is even larger than for protons. We do not plot results for 3\,TeV electrons, because their energy losses are small even at $t=10$\,kyr. Their distributions are not very different from those for protons, see upper left panel.
Finally, we note that the radial distributions $n(r)$ for electrons shown in
Ref.~\cite{Kistler:2012ag} agree approximately with ours around the peak
but lack the tails at large distances $r$.

\begin{figure}
\begin{center}
\includegraphics[width=0.33\textwidth,angle=-90]{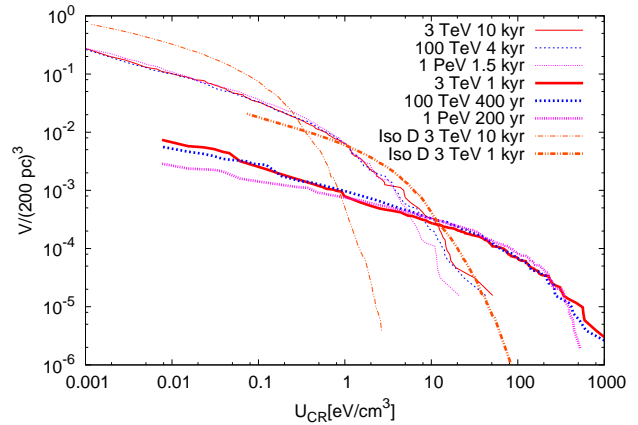}
\end{center}
\caption{Volume filling factor $f_V(>U_{\rm CR})$ for a 200\,pc cube around the source as a function of the CR energy density $U_{\rm CR}$.}
\label{VolFillingFact}
\end{figure}

As a next step, we quantify the relative sizes of regions with over-densities of CRs. 
We define the volume filling factor $f_V(>U_{\rm CR})$ as the fraction of a specified 
volume $V$ filled with CR energy densities $U_{\rm CR}(r)$ larger than a threshold 
value $U_{\rm CR}$. We choose for the volume a cube $V=(200\,{\rm pc})^3$ centered on the source 
and divide it into a large number $N$ of cells. We study the dependence of the 
filling factor on the cell size by varying it from 20 to 2.5\,pc, and then 
extrapolate to find its asymptotic value for $N\to\infty$. 
We have verified the convergence of this procedure at $E=1$\,PeV with a ten times larger 
set of CRs, $N_{\rm CR}=10^5$.

In order to determine the absolute value of the CR density, we fix the 
total energy in protons to $10^{50}$\,erg and assume a CR energy spectrum $\propto E^{-2}$. 
We divide the spectrum into 10 logarithmically spaced energy bins, thus the energy in each bin is $10^{49}$\,erg. 
We show in Fig.~\ref{VolFillingFact} the resulting volume 
filling factor as a function of the CR density for 3\,TeV protons at 
two different times, $t=1$\,kyr (thick red line) and 10\,kyr (thin red one). 
For comparison, we plot with orange dashed-dotted lines the volume filling factor expected from 
standard isotropic diffusion, according to Eq.~(\ref{n_3D}). While red and orange lines 
have in both cases approximately the same slope at low densities, the 
orange ones start to steepen earlier. This shows that for anisotropic diffusion, 
a few regions with significantly higher CR densities exist, even at $t=10$\,kyr when no filaments are visible. 
The spatial distribution is still somewhat clumpy at $t=10$\,kyr. 
Also, the fact that the orange lines are above the red ones 
at low densities $U_{\rm CR}$ points out that the bulk of CRs actually 
occupy a smaller total volume 
than one would expect for isotropic diffusion. 
While this behavior is qualitatively the same for the two times, the relative discrepancy 
between our results and isotropic ones clearly shrinks from $t=1$\,kyr to $t=10$\,kyr. 
This is in line with our results converging towards the isotropic predictions at sufficiently late times.

Another important result is presented in Fig.~\ref{VolFillingFact}. 
The volume filling factors for high energies are similar to those for 
lower energies but at earlier times: The lines for 
$\{E=3\,{\rm TeV};\,t=10\,{\rm kyr}\}$, 
$\{E=100\,{\rm TeV};\,t=4.5\,{\rm kyr}\}$, and 
$\{E=1\,{\rm PeV};\,t=1.5\,{\rm kyr}\}$ are nearly identical. 
In the same way $f_V(>U_{\rm CR})$ is similar for 
$\{E=3\,{\rm TeV};\,t=1\,{\rm kyr}\}$, 
$\{E=100\,{\rm TeV};\,t=400\,{\rm yr}\}$, and 
$\{E=1\,{\rm PeV};\,t=200\,{\rm yr}\}$. 
Therefore, our results evolve in 
a self-similar way too, as is the case with isotropic diffusion. 
If one rescales time as $t \sim E^{-1/3}$, the volume filling factor
remains invariant. Thus the volume filling factor $f_V(>U_{\rm CR})$ scales 
with the parameters energy and time as expected from isotropic diffusion, 
where $\lambda^{1/2}\sim E^{1/3}t$, while its behavior as a function of 
$U_{\rm CR}$ strongly differs from the isotropic expectation.
This scaling behavior can be used to study diffusion at low 
energies while performing calculations at higher energies but earlier times.

\section{Energy spectra of gamma-ray emissions}
\label{GRSpectra}

\begin{figure}
\begin{center}
\includegraphics[width=0.49\textwidth]{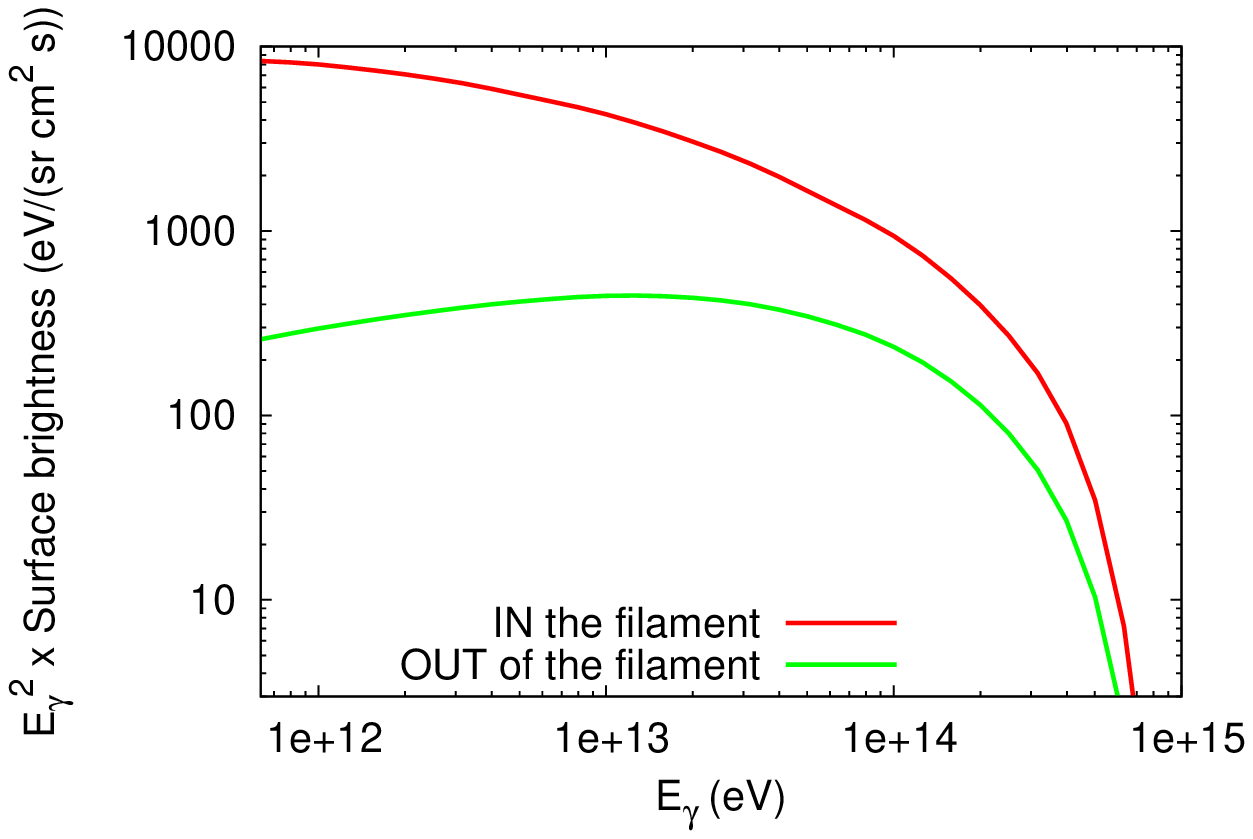}
\includegraphics[width=0.49\textwidth]{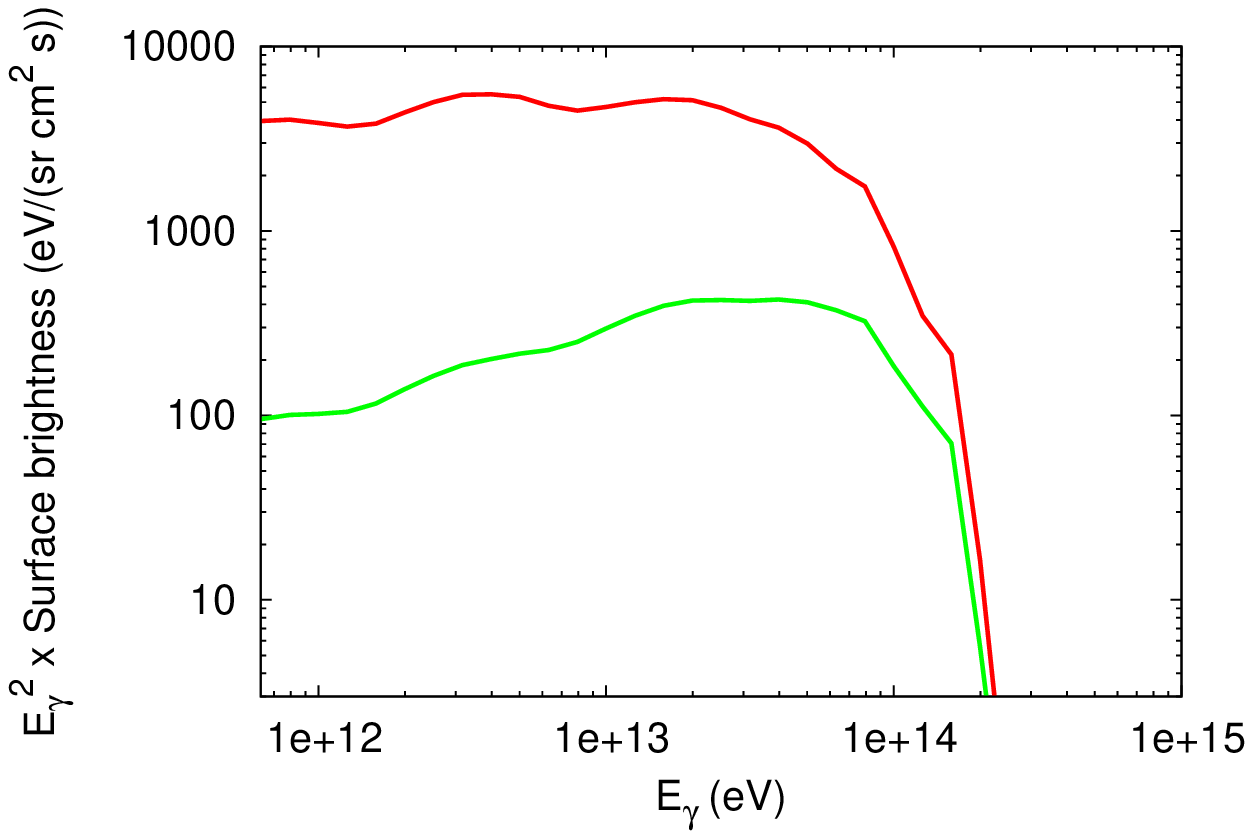}
\end{center}
\caption{Surface brightness in gamma-rays $\times E_{\gamma}^{2}$ versus gamma-ray energy at $t=2$\,kyr, for regions inside and outside a filament, for protons ({\it upper panel}) and electrons ({\it lower panel}). See text for the definition of the filament. Same parameters are used for the source as in Fig.~\ref{Filaments}. The source is located $D=3$\,kpc from the observer, and $E_{\rm CR}=10^{50}$\,erg is assumed to be given to CR protons. $n_{\rm H}=1\,$cm$^{-3}$ is used for the density of the interstellar medium. Electron flux is suppressed by $K_{\rm ep}\sim 0.3\%$ relative to the proton flux.}
\label{MC_InOut_Filaments}
\end{figure}

Lower energy CRs are expected to remain for longer times in filaments than higher energy ones, see the previous section. This implies that darker regions that surround filaments contain a higher ratio of high to low energy CRs than the filaments themselves. Therefore, the gamma-ray spectrum of the emission from filaments is steeper than that from surrounding regions. Of course, this result only holds for gamma-ray energies above which corresponding CR primaries have already escaped from the accelerator. Also, it is valid for cases where the emission from these darker surrounding regions is still dominated by CRs from the source. The difference of slope can be clearly seen in Fig.~\ref{MC_InOut_Filaments}. The red and green lines represent the emission spectra from respectively inside and outside filaments around one given source. This effect is valid both for protons (upper panel) and electrons (lower panel). In order to compute these spectra, we have taken a 2\,kyr old source displaying a clear filament, similar to those in Figs.~\ref{Filaments} and~\ref{EmissionOnTheSky}. We have computed the surface brightness in gamma-rays above $E_{\gamma} \geq 600$\,GeV, and defined the filament as the narrow and bright stripe having a surface brightness larger than a given value. The spectra in Fig.~\ref{MC_InOut_Filaments} correspond to those inside and outside this region. For the surface brightness normalization, we have assumed that $10^{50}$\,erg has been channeled into CRs, and that the electron flux is suppressed by a factor $K_{\rm ep}\sim 0.3\%$ relative to the proton flux. The source is located at a distance $D=3$\,kpc from Earth and the thermal proton density in the surrounding interstellar medium is assumed to be $n_{\rm H}=1\,$cm$^{-3}$.

Molecular clouds (MCs) around sources start to shine brightly in gamma-rays once hadronic CRs released by the accelerator reach them. In the following, we will not consider clouds directly interacting with supernova shocks or their precursors. We only consider clouds that are further away, i.e., those that probe CRs that have already escaped from the source. One notably uses MCs to derive information on CR sources (CR acceleration and escape) and to derive the CR diffusion coefficient around them. Isotropic diffusion is an underlying assumption of the vast majority of works on this subject~\cite{Gabici:2010iw,Gabici:2012sp}, with the notable exceptions of Refs.~\cite{Giacinti:2012ar,Malkov:2012qd,Nava:2012ga,Gabici:2013eha}.

However, we have shown that in some realizations of the surrounding interstellar magnetic fields, one could expect differences of more than two orders of magnitude between the extreme eigenvalues of the CR diffusion tensor~\cite{Giacinti:2012ar}. On average, this ratio is found to be at least of a few tens for a Kolmogorov spectrum and $l_{\max}=150$\,pc. This effect is maximal at ``early times'' after CR release, typically for $t \lesssim t_{\ast}/10$, with $t_{\ast}$ defined as in Eq.~(\ref{tast}). This has crucial implications for studies of gamma-ray emission from MCs surrounding CR sources and located within ${\cal O}(l_{\max})$ from them. Our work calls for a careful analysis of such data. For instance, the CR diffusion coefficient value in the interstellar medium cannot be estimated from only a few clouds.

In addition to this, we have found a new type of spectrum distortion due to the difference between gamma-ray spectra from inside and outside filaments. Usually, the gamma-ray spectra from MCs are assumed to differ from that of the parent CR distribution accelerated at the source due to two effects. First, a distortion arises from the fact that CRs with different energies escape from their sources at different times. Second, the CR propagation times from the source to the cloud depend on CR energies, resulting in another distortion. Our results in Fig.~\ref{MC_InOut_Filaments} imply that the gamma-ray spectra from molecular clouds located in filaments should be expected to be steeper than those from clouds located outside. In practice, clouds aligned with filaments should be preferentially detected because they are significantly brighter than those outside. Therefore, a selection effect may be at work: Clouds that are preferentially detected in gamma-rays display a steeper spectrum than what would be expected if CRs were to diffuse isotropically around their sources. This third distortion should be added to the two previous ones. Indeed, at a given time of observation, clouds located at identical radii around a given source should have identical spectra if only the first two distortion effects were present. Nevertheless, they have steeper or flatter spectra depending on whether or not they are aligned with a CR filament.

From one source to another, the given realization of the surrounding interstellar magnetic fields results in a wide variety of possible shapes for the CR distribution. While the above effects complicate the analysis of gamma-ray emissions from molecular clouds around sources, they also bring new and interesting perspectives to this field. In principle, information on the local magnetic fields may be retrieved from such data in the future, as discussed in Section~\ref{LocalMF}.

\begin{figure*}[!t]
  \centerline{\includegraphics[width=0.33\textwidth]{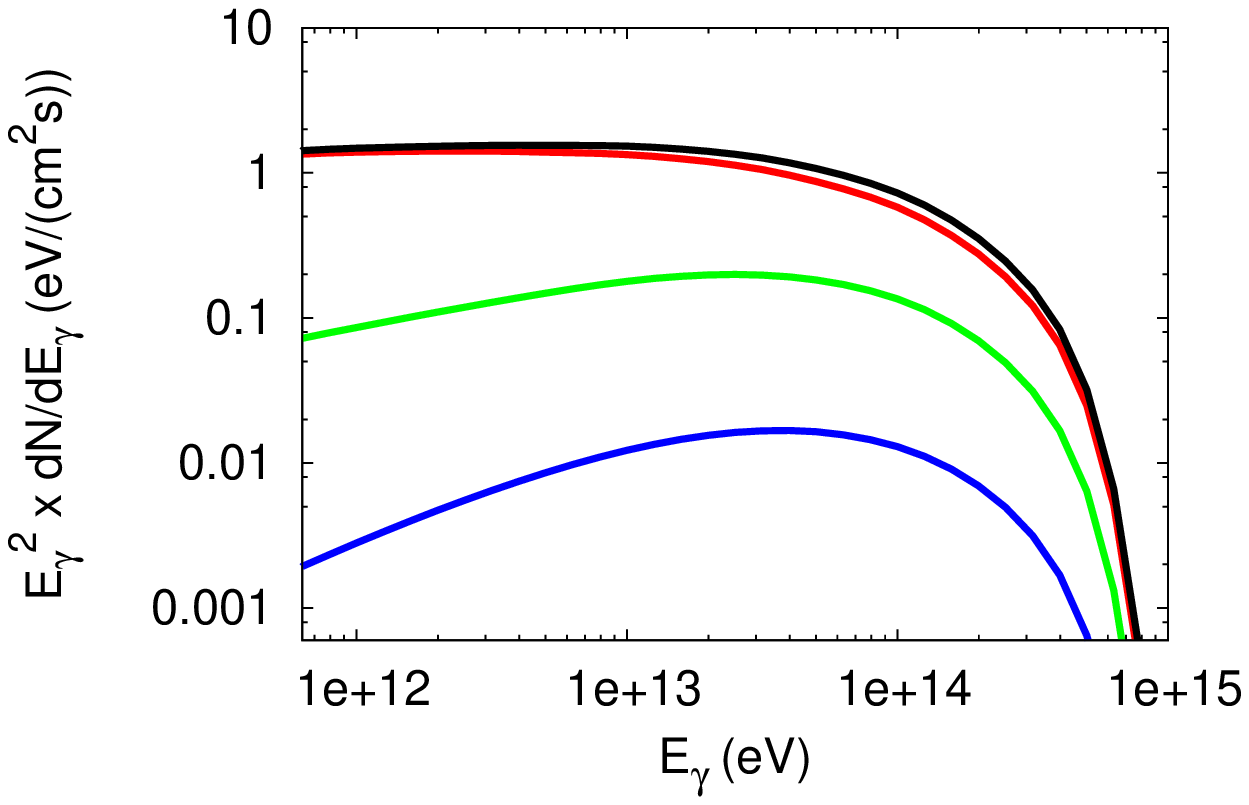}
              \hfil
              \includegraphics[width=0.33\textwidth]{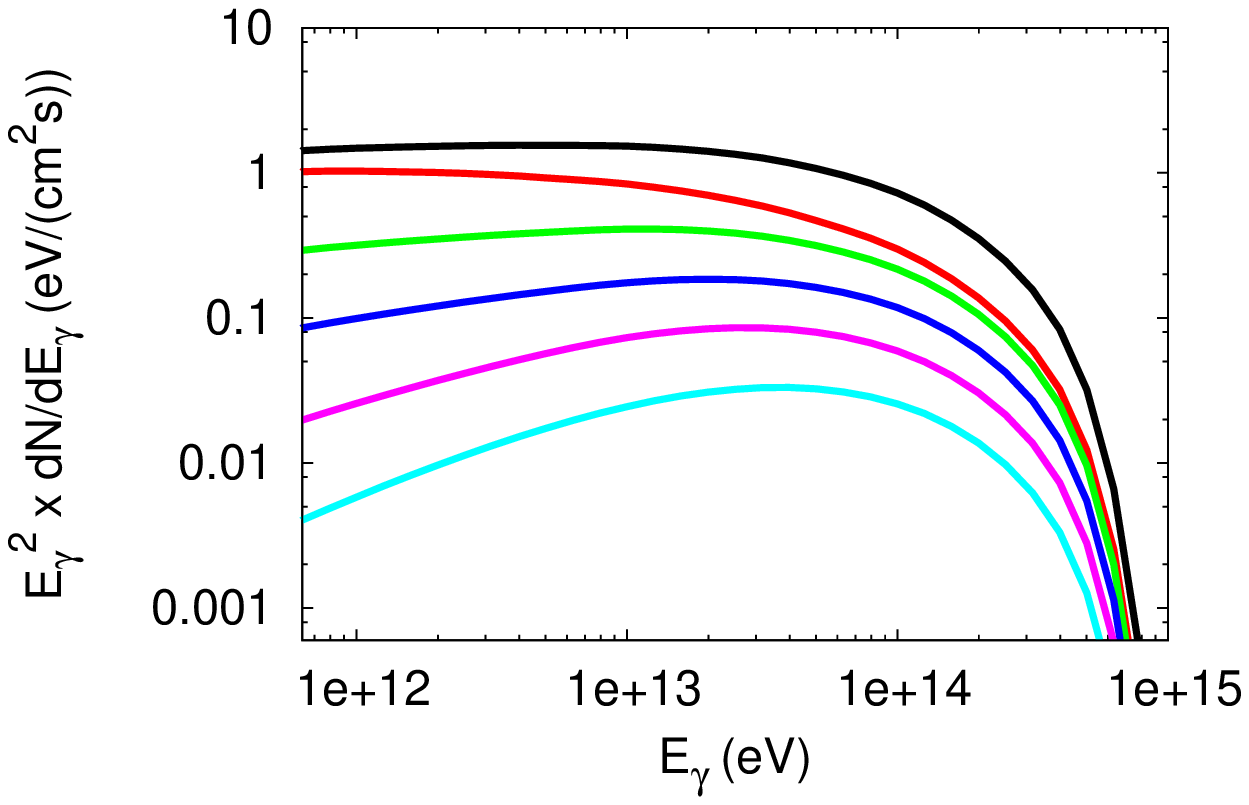}
              \hfil
              \includegraphics[width=0.33\textwidth]{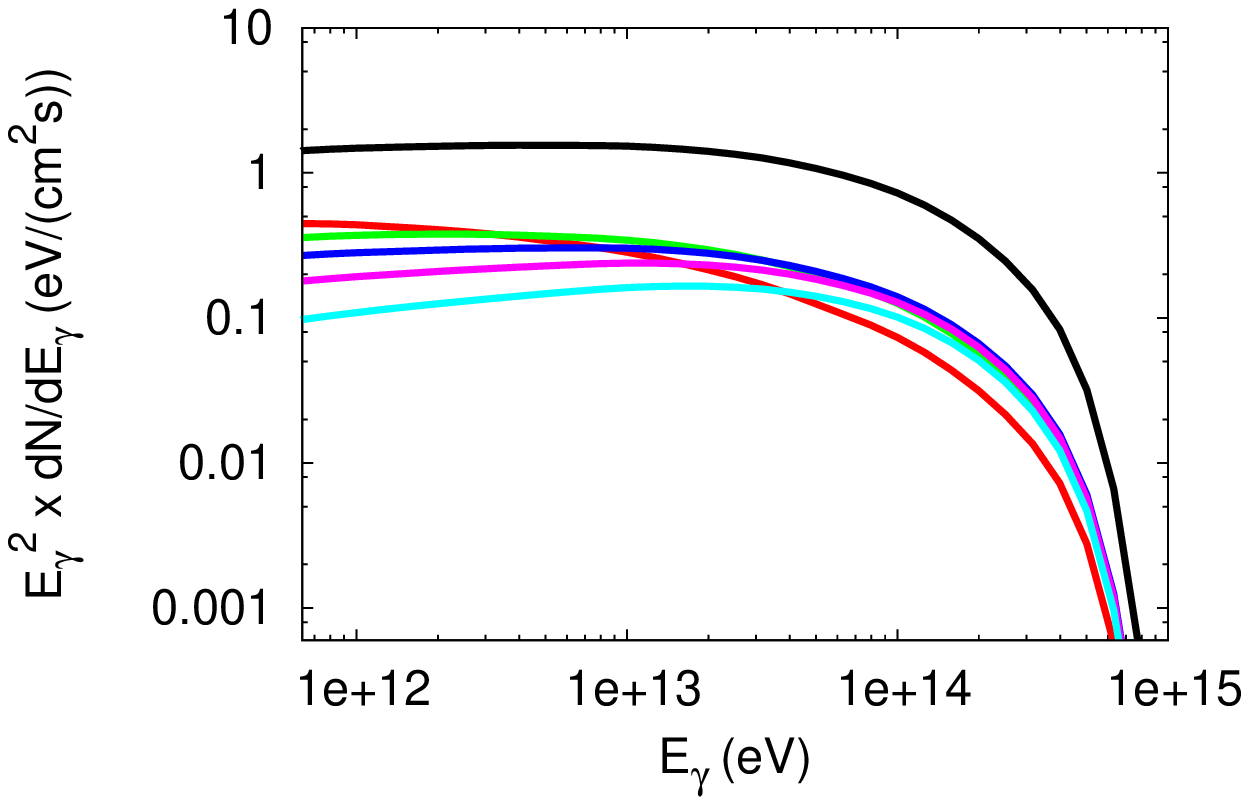}
             }
  \centerline{\includegraphics[width=0.33\textwidth]{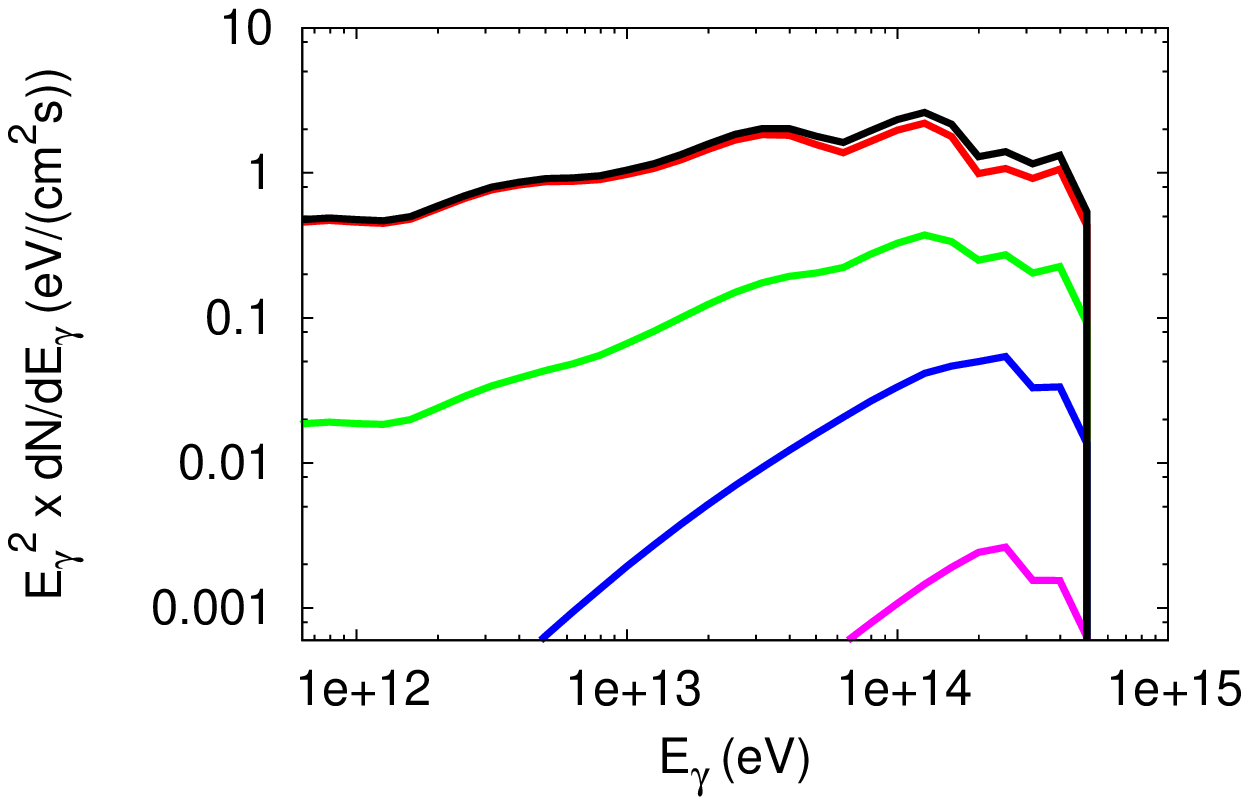}
              \hfil
              \includegraphics[width=0.33\textwidth]{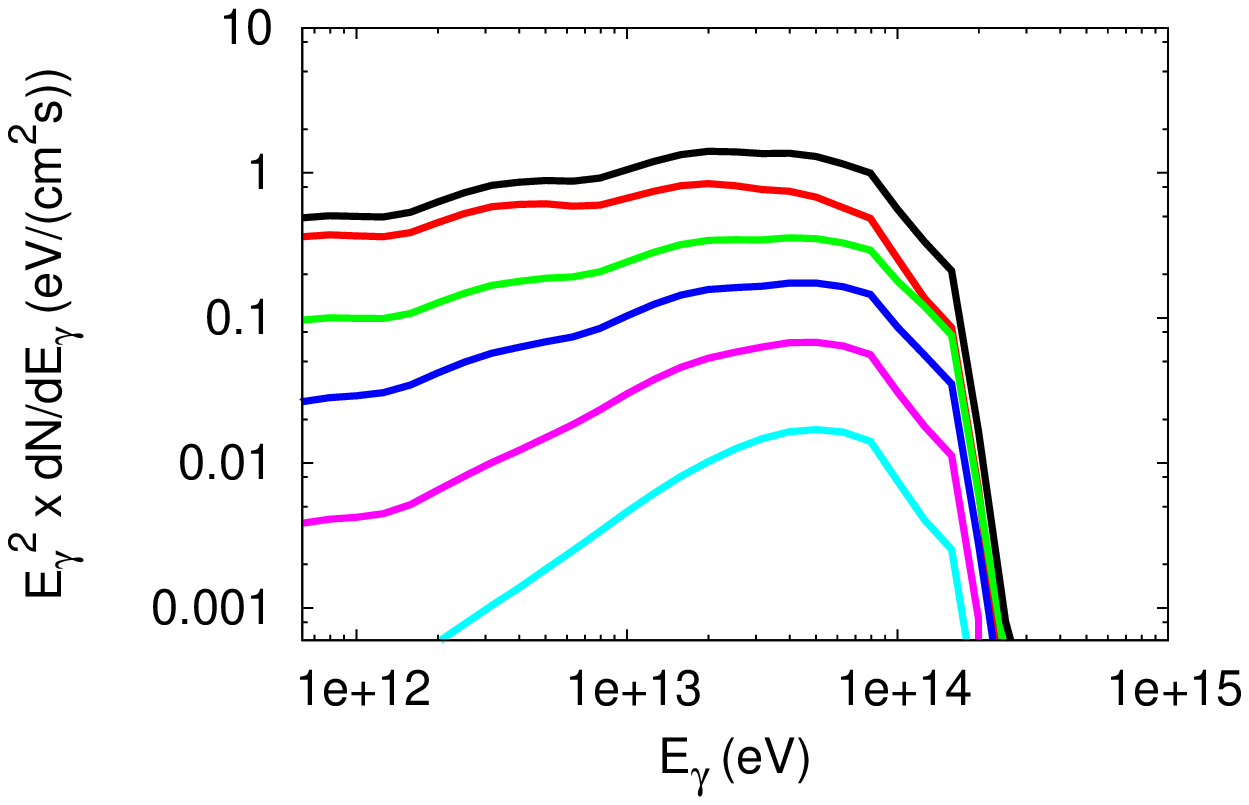}
              \hfil
              \includegraphics[width=0.33\textwidth]{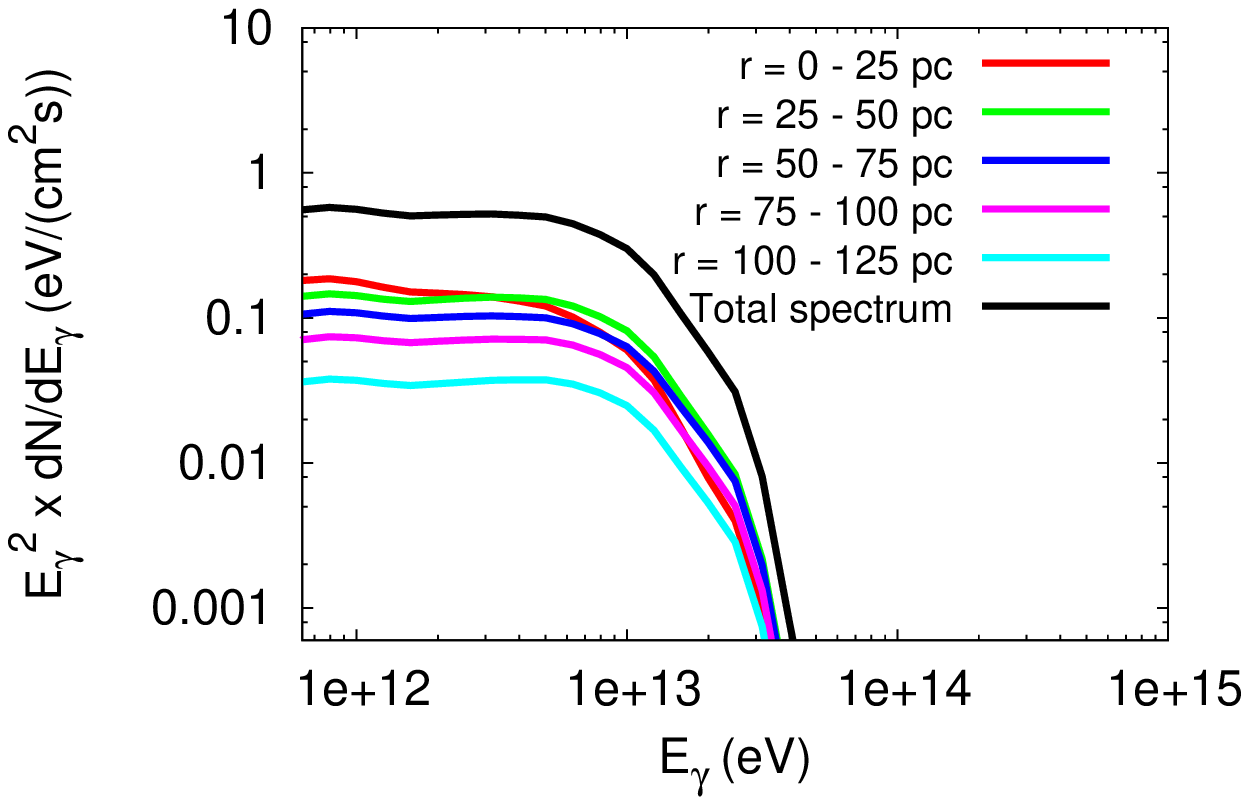}
             }
  \caption{Gamma-ray spectra in 2D concentric tori centered on the source for a proton ({\it upper row}) and an electron ({\it lower row}) source at times $t=0.5,\,2,\,10$\,kyr ({\it resp. left, middle and right columns}). See key for radii of tori on the lower left panel. Same parameters are used as in Fig.~\ref{MC_InOut_Filaments}.}
  \label{GammaRaySpectra}
\end{figure*}

For the same reason, if one divides the sky around sources into four quadrants (or more generally into $\mathcal{N}$ circular sectors), the gamma-ray spectrum of the associated extended gamma-ray emission is expected to differ from one quadrant (or circular sector) to another. We have checked this and found that in some cases results differ significantly from one quadrant to another, despite the fact that all quadrants encompass the same range of distances from the source. In order to analyze the differences in the spectra from one region of the extended emission to another, one may also divide the sky with concentric tori centered on the source. Such a study is still rather theoretical given the present angular resolution and amount of data, but it will become easier in the future. Figure~\ref{GammaRaySpectra} presents such spectra, for five concentric tori encompassing respectively the following ranges of radii: $r = 0$--25, 25--50, 50--75, 75--100 and 100--125\,pc. Three times after CR escape are considered $t=0.5,\,2,\,10$\,kyr, for respectively the left, middle and right columns. The first row shows gamma-ray spectra for proton primaries and the second one for electron primaries. The relative normalizations for the fluxes $E_{\gamma}^{2} \times dN/dE_{\gamma}$ have been made using the same parameters as for Fig.~\ref{MC_InOut_Filaments}. The shapes of the spectra would not change for other values of parameters $E_{\rm CR},\,K_{\rm ep},\,D$ and $n_{\rm H}$, and other normalizations can be easily deduced from ours. In practice, such curves should still be convolved with a given template for the energy-dependent escape of CRs from the source, see Section~\ref{Code} for a discussion. The black lines represent the total spectrum of the emission from the whole field-of-view. The spectrum for electrons changes in time because of energy losses. When assuming identical electron and proton spectral indexes at $t=0$, it starts to look as steep as that from protons at $t=10$\,kyr, in the considered photon energy range $E_{\gamma}=600$\,GeV -- 100\,TeV. At lower energies $E_{\gamma}$, a difference between electrons and protons would still be visible: There is no significant pileup of high energy electrons to lower energies on such time scales. As expected, the spectrum of the emission from outer tori looks much flatter than the average one. The latter is about $\propto E_{\gamma}^{-2}$ for proton primaries following an $E^{-2}$ spectrum. On the contrary, the spectrum of the central torus becomes steeper, especially at later times. As should be expected, these differences of slopes are due to the fact that, globally, higher energy CRs still diffuse faster from the source than lower energy ones.

\section{Conclusions and perspectives}
\label{Conclusions}

In this work, we have pointed out a significant inconsistency between the present knowledge of the interstellar magnetic field and the assumptions used to describe CR diffusion on scales $\lesssim {\cal O}(100\,$pc). Cosmic ray diffusion around sources and local CR propagation within $\sim {\rm few}\times 100$\,pc from Earth are affected. These findings put together with those of Ref.~\cite{Giacinti:2011mz} allow us to get a self-consistent explanation for CR anisotropies observed at large and small scales on the sky by IceCube~\cite{ICobs} and other observatories~\cite{CRAobs}. We have investigated here the impacts on gamma-ray astronomy of Galactic sources. Our findings call for a shift in the standard assumptions used in this field. Presently, the vast majority of studies of gamma-rays from molecular clouds or from extended emissions around sources are based on the assumption of isotropic CR diffusion. The impact of the regular Galactic magnetic field is sometimes added. We have shown that none of these descriptions are close to being satisfactory and that they can lead to wrong conclusions when analyzing recent sources. Even in isotropic magnetic turbulence, large scale modes act as local regular fields and drive a strongly anisotropic diffusion of CRs. Within a box of size $\sim l_{\max}^{3}$, the direction of modes with scales $\gg r_{\rm L}$ varies in space, which leads to twisted and irregular shapes of the CR distributions around sources.

Figures~\ref{Filaments} and~\ref{EmissionOnTheSky} present simulated images of sources in gamma-rays and illustrate some of the challenges for gamma-ray astronomy. The extended emissions around sources are not always centered on the sources, which may lead to misidentifications. For the sake of clarity, an isotropic density of thermal protons has been assumed. Even so, images can display counter-intuitive features, such as multiple and curved 'filaments'. In some cases, surface brightness may locally increase with distance from the source.

We have proposed in Section~\ref{LocalMF} that gamma-ray astronomy can be used in the future as a new way to probe the interstellar turbulent magnetic fields. It will enable one to get independent information on local fields around CR sources and to compare it to other observations, such as unpolarized synchrotron data. By analyzing several sources, one can get a precious insight into parameters of the turbulence, such as $l_{\rm c}$ and $l_{\max}$. Details of the CR distribution around sources strongly vary from one source to another. Very schematically, we would expect that for $t \lesssim t_{\ast}/10$ after release, with $t_{\ast}$ defined as in Eq.~(\ref{tast}), CRs are mainly confined in a filamentary flux tube containing the source. Its degree of collimation depends on the local field realization. Information on the field strength and direction may be retrieved from observations. For $t_{\ast}/10 \lesssim t \lesssim t_{\ast}$, the distribution is less anisotropic, though still significantly different from what one would expect if diffusion were isotropic. Gamma-ray data can then notably be used to estimate $l_{\max}$.

Our results converge towards the predictions of the standard CR diffusion approximation on scales $\gtrsim {\cal O}(l_{\max})$. However, a non-negligible difference still exists within a few $l_{\max}$ from the source in the large distance tail of the CR distribution, see Section~\ref{AnisoDiff}.

In Section~\ref{GRSpectra}, we have computed spectra from different regions of extended gamma-ray emissions, and from molecular clouds around sources. We have found an additional distortion of the energy spectra of gamma-rays from clouds, on top of those due to an energy-dependent escape from the source and an energy-dependent diffusion coefficient. We have shown that for identical MCs lit by CRs streaming from the same source, the brightest clouds would have steeper spectra than dimmer ones.

In the future, better angular resolution of gamma-ray observatories, as well as larger statistics, will allow one to investigate anisotropic diffusion of CRs around sources in more detail. If systematically no anisotropy or only very little were to be found, this would have important implications, notably on our knowledge of the turbulent Galactic magnetic field. Likely reasons for very little anisotropy would be either a small coherence length of the turbulence (significantly smaller than a few tens of parsecs), or a lack of power in its large scale modes, compared to Kraichnan and Kolmogorov spectra for example. However, we note that first hints in favor of our findings may be seen in the irregular gamma-ray halos observed around CR proton and electron sources~\cite{NS12,HESS,VERITAS}. Our results could also explain the offsets observed between some gamma-ray emissions and their most likely sources, such as that between HESS J1303-631 and PSR 1301-6305~\cite{Aharonian:2005rv,Abramowski:2012fc}. In addition, the recent work of Nava and Gabici~\cite{Nava:2012ga} may hint at an anisotropic CR distribution around the supernova remnant W28. In both cases, gamma-ray observatories will be able to provide valuable information on the structure of interstellar magnetic fields around sources.

\section*{Acknowledgments}

We thank Tony Bell, Stefano Gabici and Brian Reville for useful discussions. 
The authors acknowledge the support provided by the Fran\c{c}ois Arago Centre 
at the APC laboratory in Paris.
GG acknowledges funding from the European Research Council under the European Community's Seventh Framework Programme (FP7/$2007-2013$)/ERC grant agreement no. 247039.

\appendix
\section{CR-driven instabilities and interstellar magnetic fields around sources}
\label{App}

In this appendix, we briefly quantify and discuss the limits of our test particle approach. The extent to which CRs that have escaped from their sources modify interstellar magnetic fields on the length scales considered here depends on several unknown parameters. We show below that even in the 'worst case' scenario, one should not expect significant deviations from our test particle case for CRs with $E \gtrsim$\,a few tens of TeV, i.e., $E_{\gamma} \gtrsim$\,a few TeVs. Deviations are expected in regions with large CR currents $j_{\rm CR}(E)$, especially close to the source, at early times after release and within strongly collimated CR filaments. Once CRs have left filaments, the CR current $j_{\rm CR}$ drops sharply. Assuming an $E^{-2}$ spectrum, $10^{50}$\,erg in CRs from 1\,GeV to 1\,PeV and ten energy bands with logarithmic widths, $10^{49}$\,erg is contained in each band. As discussed in Section~\ref{LocalMF}, escaping CRs are initially more or less focused in a magnetic flux tube containing the source. In a very well collimated CR filament of radius $\approx 3$\,pc and length $\approx 2 l_{\rm c} \approx 60$\,pc, the CR energy density reaches $U_{\rm CR}\sim 100$\,eV$\cdot$cm$^{-3}$. We assume that CRs propagate in the filament with a speed $\sim D/l_{\rm c}$ where $D=D_{0}E^{1/3}$ is the CR parallel diffusion coefficient. Then, $j_{\rm CR} \sim U_{\rm CR}eD/(El_{\rm c})$. Here, $D_{0} \sim 10^{29}$\,cm$^{2}\,$s$^{-1}\,$PeV$^{-1/3}$ from the largest eigenvalue in Fig.~1 of Ref.~\cite{Giacinti:2012ar}. Bell's non-resonant hybrid (BNRH) instability~\cite{Bell2004} dominates over the Alfv\'en instability if $Bj_{\rm CR}r_{\rm L}/(\rho_{\rm ISM}v_{\rm A}^{2})>1$. Using $\rho_{\rm ISM} \approx 1\,m_{\rm p}\,$cm$^{-3}$ as the density of the interstellar medium (ISM) and $B\simeq4\,\mu$G, we find that $Bj_{\rm CR}r_{\rm L}/(\rho_{\rm ISM}v_{\rm A}^{2}) \sim 1$. Our conclusions do not change much with the dominating instability: Their respective growth rates are $\Gamma_{\rm BNRH} = 0.5j_{\rm CR}\sqrt{\mu_0/\rho_{\rm ISM}}$ and $\Gamma_{\rm Alf} \approx 0.3j_{\rm CR}\sqrt{\mu_0/\rho_{\rm ISM}}$, as can be deduced from the dispersion relations~(7ff) of Ref.~\cite{Bell2004}. With the above parameters, the typical growth time $\approx 5 \Gamma_{\rm BNRH}^{-1} \approx 10\sqrt{\rho_{\rm ISM}/\mu_0}E^{2/3}l_{\rm c}/(U_{\rm CR}eD_0)$ is $\approx 1.4$, 3.1, 14 and 67\,kyr for respectively 3, 10, 100 and 1000\,TeV CRs. Estimating roughly $t_{\rm c} \sim l_{\rm c}^{2}/D$ as the time spent in a collimated filament by the bulk of escaping CRs, instabilities do not have enough time to grow significantly for $E \gtrsim 0.1 \sqrt{\mu_0/\rho_{\rm ISM}}U_{\rm CR}el_{\rm c} \approx 40$\,TeV even in well collimated filaments. In practice, this value would be further diminished if the damping rate of waves is large enough. For instance, in a region with $10^{-2}$ neutral hydrogen atoms per cm$^{3}$ and $T=10^{3}$\,K, the ion-neutral damping rate is $\approx 1\,$kyr$^{-1}$~\cite{KC1971,Bell1978}, which would lower the above limit by a factor of a few. See for example Refs.~\cite{Damping} for other sources of damping. A full treatment is beyond the scope of the present work, but we expect that if and when CR back-reaction is non-negligible, the diffusion coefficient along the filament would be suppressed and CRs would diffuse more isotropically---perpendicular to the filament. This would in turn lower $U_{\rm CR}$ and $\Gamma_{\rm BNRH,Alf}$ (very roughly by a factor $\sim 10$ for a factor 3 increase in the radius of the filament), which suggests that some anisotropies would remain at $E=3$\,TeV even in such a case. Let us also note that, within the time scale $t_{10k}$ we consider, the CR pressure inside filaments is not sufficient to significantly expand them laterally. An ISM fluid parcel on the side would be accelerated as $\rho_{\rm ISM}du/dt = |{\bf \nabla}P_{\rm CR}| \sim 100\,$eV$\cdot$cm$^{-3}/$3\,pc. This leads to a displacement $\sim t_{10k}^{2}du/dt \approx 0.3$\,pc, which is smaller than the width of filaments.


\end{document}